\documentclass[twoside,twocolumn,9pt]{article}
\usepackage{extsizes}
\usepackage[super,sort&compress,comma]{natbib} 
\usepackage[version=3]{mhchem}
\usepackage[left=1.5cm, right=1.5cm, top=1.785cm, bottom=2.0cm]{geometry}
\usepackage{balance}
\usepackage{mathptmx}
\usepackage{amssymb}
\usepackage{sectsty}
\usepackage{graphicx} 
\usepackage{lastpage}
\usepackage[format=plain,justification=justified,singlelinecheck=false,font={stretch=1.125,small,sf},labelfont=bf,labelsep=space]{caption}
\usepackage{float}
\usepackage{fancyhdr}
\usepackage{fnpos}
\usepackage[english]{babel}
\addto{\captionsenglish}{%
  
}
\usepackage{array}
\usepackage{droidsans}
\usepackage{charter}
\usepackage[T1]{fontenc}
\usepackage[usenames,dvipsnames]{xcolor}
\usepackage{setspace}
\usepackage[compact]{titlesec}
\usepackage{hyperref}
\usepackage{comment}
\usepackage[table]{xcolor}
\usepackage{epstopdf}%This line makes .eps figures into .pdf - please comment out if not required.

\definecolor{cream}{RGB}{222,217,201}

\newcommand{\torem}[1]{\textcolor{gray}{#1}}

\begin{document}

\pagestyle{fancy}
\thispagestyle{plain}
\fancypagestyle{plain}{
%%%HEADER%%%
\renewcommand{\headrulewidth}{0pt}
}
%%%END OF HEADER%%%

%%%PAGE SETUP - Please do not change any commands within this section%%%
\makeFNbottom
\makeatletter
\renewcommand\LARGE{\@setfontsize\LARGE{15pt}{17}}
\renewcommand\Large{\@setfontsize\Large{12pt}{14}}
\renewcommand\large{\@setfontsize\large{10pt}{12}}
\renewcommand\footnotesize{\@setfontsize\footnotesize{7pt}{10}}
\makeatother

\renewcommand{\thefootnote}{\fnsymbol{footnote}}
\renewcommand\footnoterule{\vspace*{1pt}% 
\color{cream}\hrule width 3.5in height 0.4pt \color{black}\vspace*{5pt}} 
\setcounter{secnumdepth}{5}

\makeatletter 
\renewcommand\@biblabel[1]{#1}            
\renewcommand\@makefntext[1]% 
{\noindent\makebox[0pt][r]{\@thefnmark\,}#1}
\makeatother 
\renewcommand{\figurename}{\small{Fig.}~}
\sectionfont{\sffamily\Large}
\subsectionfont{\normalsize}
\subsubsectionfont{\bf}
\setstretch{1.125} %In particular, please do not alter this line.
\setlength{\skip\footins}{0.8cm}
\setlength{\footnotesep}{0.25cm}
\setlength{\jot}{10pt}
\titlespacing*{\section}{0pt}{4pt}{4pt}
\titlespacing*{\subsection}{0pt}{15pt}{1pt}
%%%END OF PAGE SETUP%%%

%%%FOOTER%%%
\fancyfoot{}
\fancyfoot[LO,RE]{\footnotesize\sffamily Preprint submitted to \textit{Soft Matter}}
\fancyfoot[RO,LE]{\footnotesize\sffamily \thepage~\textbar~\pageref{LastPage}}
%\fancyfoot[LO,RE]{\vspace{-7.1pt}\includegraphics[height=9pt]{head_foot/LF}}
%\fancyfoot[CO]{\vspace{-7.1pt}\hspace{13.2cm}\includegraphics{head_foot/RF}}
%\fancyfoot[CE]{\vspace{-7.2pt}\hspace{-14.2cm}\includegraphics{head_foot/RF}}
%\fancyfoot[RO]{\footnotesize{\sffamily{1--\pageref{LastPage} ~\textbar  \hspace{2pt}\thepage}}}
%\fancyfoot[LE]{\footnotesize{\sffamily{\thepage~\textbar\hspace{3.45cm} 1--\pageref{LastPage}}}}
\fancyhead{}
\renewcommand{\headrulewidth}{0pt} 
\renewcommand{\footrulewidth}{0pt}
\setlength{\arrayrulewidth}{1pt}
\setlength{\columnsep}{6.5mm}
\setlength\bibsep{1pt}
%%%END OF FOOTER%%%

%%%FIGURE SETUP - please do not change any commands within this section%%%
\makeatletter 
\newlength{\figrulesep} 
\setlength{\figrulesep}{0.5\textfloatsep} 

\newcommand{\topfigrule}{\vspace*{-1pt}% 
\noindent{\color{cream}\rule[-\figrulesep]{\columnwidth}{1.5pt}} }

\newcommand{\botfigrule}{\vspace*{-2pt}% 
\noindent{\color{cream}\rule[\figrulesep]{\columnwidth}{1.5pt}} }

\newcommand{\dblfigrule}{\vspace*{-1pt}% 
\noindent{\color{cream}\rule[-\figrulesep]{\textwidth}{1.5pt}} }

\makeatother
%%%END OF FIGURE SETUP%%%

%%%TITLE, AUTHORS AND ABSTRACT%%%
\twocolumn[
  \begin{@twocolumnfalse}
{\includegraphics[height=30pt]{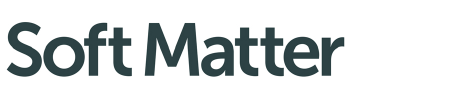}\hfill\raisebox{0pt}[0pt][0pt]{\includegraphics[height=55pt]{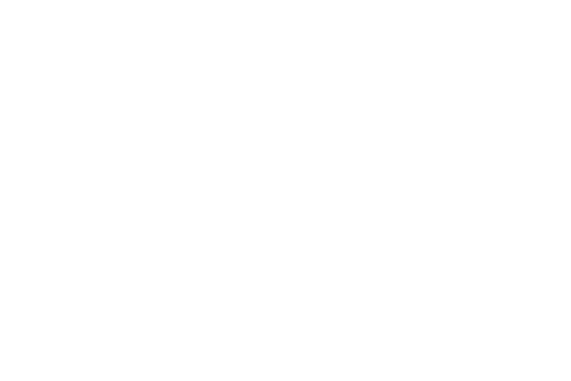}}\\[1ex]
\includegraphics[width=18.5cm]{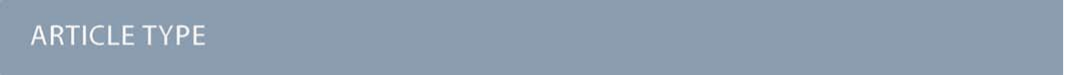}}\par
\vspace{1em}
\sffamily
\begin{tabular}{m{4.5cm} p{13.5cm} }

\includegraphics{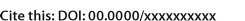} & \noindent\LARGE{\textbf{Indentation of pre-tensioned cortex-like fibre networks: tension, local stretching and the signature of defect architecture}} \\%Article title goes here instead of the text "This is the title"
\vspace{0.3cm} & \vspace{0.3cm} \\

 & \noindent\large{Shubham Agarwal$^{\ast}$\textit{$^{a}$}, Olivier No\"el\textit{$^{b}$}, Martin H. M\"user\textit{$^{a}$}} \\%Author names go here instead of "Full name", etc.

\includegraphics{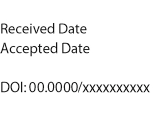} & \noindent\normalsize{
Indentation is widely used to infer the mechanical state of biological materials like the actomyosin cortex. 
It reduces a single force–indentation curve to an apparent modulus. 
The cortex, however, is a pre-tensioned, heterogeneous fibre network.
We show that this inference is ambiguous. 
The same amount of weakening can leave the indentation force unchanged or lower it tenfold, depending on where it lies relative to the probe. 
The force reflects how the probe couples to the tensioned, load-bearing structure of the layer, not the amount of weakening.
We combine molecular dynamics simulations of a pressure-loaded, crosslinked fibre network indented by a rigid sphere with a continuum membrane solution.
We represent the simulated force--depth response by a two-term fit, $F\propto a\delta+b\delta^{3/2}$, with $a$ reflecting the global tension, and $b$ the stiffening with depth.
%
%A membrane solution for a sphere indenting the pressurised layer at this tension, with no adjustable parameter, predicts the linear coefficient $a$ of the healthy network to within 3\% and accounts for half of $b$; letting the tension rise near the contact due to local stretching, recovers most of the remainder.
%
We investigate two forms of weakening: diffuse and clustered.
At a fixed fraction of weakened bonds, diffuse weakening lowers the force by a factor of about 1.5.
A weak cluster beneath the probe lowers it 8–12-fold, because the slack cluster decouples the probe from the tension.
When the clusters are displaced from the contact, the force returns to within 15\% of the healthy value. 
Two opposite effects cancel. 
The clusters soften the layer as a whole, which lowers its tension and the force. 
But they also channel the load through the intact network beneath the probe, which stiffens it locally and raises the force.
The shape of the curve retains what the force level hides. 
The coefficient $a$ reflects the softened layer when clusters beside the probe keep the force near the healthy value.
The local exponent $\alpha=\mathrm{d}\ln F/\mathrm{d}\ln\delta$ increases with depth in every configuration except when a cluster lies beneath the probe, where the two-term form no longer applies.
Detecting localised weakening therefore requires spatial sampling and analysis of the curve shape, not a single fitted modulus.
} 
\end{tabular}

 \end{@twocolumnfalse} \vspace{0.6cm}
  ]
%%%END OF TITLE, AUTHORS AND ABSTRACT%%%

%%%FONT SETUP - please do not change any commands within this section
\renewcommand*\rmdefault{bch}\normalfont\upshape
\rmfamily
\section*{}
\vspace{-1cm}

%%%FOOTNOTES%%%

\footnotetext{\textit{$^{a}$~Address: Department of Material Science and Engineering, Saarland University, Saarbr\"ucken-66123, Saarland, Germany, $^{b}$ Address: Institut des Molécules et Matériaux du Mans, Le Mans Université, 72085 Cedex 09, Le Mans, France, $^*$ E-mail: shubham.agarwal@uni-saarland.de}}  %E-mail: shubham.agarwal@uni-saarland.de}}
%\footnotetext{\textit{$^{b}$~Address, Address, Town, Country. }}

%Please use \dag to cite the ESI in the main text of the article.
%If you article does not have ESI please remove the the \dag symbol from the title and the footnotetext below.
%\footnotetext{\dag~Supplementary Information available: [details of any supplementary information available should be included here]. See DOI: 10.1039/cXsm00000x/}
%additional addresses can be cited as above using the lower-case letters, c, d, e... If all authors are from the same address, no letter is required

%\footnotetext{\dag~Additional footnotes to the title and authors can be included \textit{e.g.}\ `Present address:' or `These authors contributed equally to this work' as above using the symbols: \ddag, \textsection, and \P. Please place the appropriate symbol next to the author's name and include a \texttt{\textbackslash footnotetext} entry in the the correct place in the list.}

%\ddag for both dagguers amd reintroduce the part about the SI
%%%END OF FOOTNOTES%%%

%%%MAIN TEXT%%%%
Fibrous networks form a broad class of soft materials, from collagen-rich tissues~\cite{Fritton2008book,Baker2007book}, plant cell walls~\cite{butchosa2019cellulose,Hepworth2001JB} and the cytoskeleton~\cite{huisman200PRL,Zagar2011MacMol} to paper~\cite{Page1985papermaking}, polymer gels~\cite{Ostoja2002AMR} and fibre-reinforced composites~\cite{Gibson_ashby1997book}.
Their mechanical response is controlled not only by the properties of individual fibres but also by network connectivity, crosslinking, excluded-volume interactions and deformation modes such as fibre bending, stretching and non-affine rearrangement~\cite{picu2022book,Agarwal2023MRC,Agarwal2024MSMSE,Dey2026IJSS,Amjad2025JMPS,Agarwal2025IJSS}. 
Depending on the relative importance of these mechanisms and of thermal fluctuations, networks behave as athermal, thermal or semiflexible systems~\cite{picu2022book}, and their response depends on internal length scales and on spatial organisation.

A biologically important example is the actomyosin cortex, a thin crosslinked layer of actin filaments, actin-binding proteins and myosin motors beneath the plasma membrane, which regulates cell shape, division, migration and mechanosensing~\cite{salbreux2012TCS,chugh2018JCS}. 
The cortex is a structured and continually remodelled network whose mechanical response depends on filament density, crosslinking and local connectivity~\cite{chugh2017Nature}. 
Nucleation, polymerisation, severing and crosslinker-mediated reorganisation make the architecture of the cortex spatially heterogeneous~\cite{fritzsche2016ScAdv}, visible for instance as protrusions and membrane ruffles~\cite{bernitt2017NatCom}.
Altered cortical organisation accompanies pathological changes in cell shape, division and invasion~\cite{chugh2018JCS,svitkina2020TCB}. 
The cortex is, moreover, under tension, generated by myosin-II contractility and intracellular pressure~\cite{stewart2011nature,fischer2014ScRep}.

Atomic force microscopy (AFM) probes such a layer by pressing a tip into the cell and recording the force as a function of depth with nanometre resolution~\cite{binnig1986PRL}.
The indentation curve is conventionally quantified by an apparent modulus or stiffness by fitting a Hertzian or Sneddon-type contact model~\cite{sneddon1965IJES}. 
This modulus is sensitive to the cortex, and shifts substantially when actin is disrupted or contractility changes~\cite{rotsch2000BJ,warmt2021NJP}.
It depends as strongly on the measurement: on indenter geometry, sample thickness, substrate and cell layering~\cite{dimitriadis2002BJ,rheinlaender2020NatMat,Wu2018Nat}, with sharp tips and micron-scale spheres differing nearly tenfold on the same tissue~\cite{Sicard2017JMBBM,VargasPinto2013BJ}.
Nor does it reliably track the state of the cell.
Malignant cells are often reported to be markedly softer than benign ones~\cite{cross2007NatNano}. 
However, this is not universal: depending on the cell type, cancer cells have been found either softer than their normal counterparts or of similar stiffness~\cite{pei2019JCanc,rodriguezramos2014BJN}, making stiffness an ambiguous marker of the state of the cell.
Moreover, tumours are mechanically heterogeneous down to the scale of single cells, with rigid and soft cells coexisting~\cite{fuhs2022NatPhy} and cells at the invasive front softer than those in the core~\cite{han2020NatCom}. 
The measured stiffness therefore depends on where the probe is placed~\cite{Plodinec2012,Wu2018Nat}.

Two features of the cortex make this mapping particularly ambiguous. 
First, the cortex is pre-tensioned. 
When a tensioned membrane is indented, its force–indentation relation is governed by the tension as well as elasticity~\cite{Sen2005BJ}. 
An analogous competition between surface tension and bulk elasticity sets the indentation response of soft solids~\cite{style2017ARCMP,hui2015PRSA}. 
A contact model for an unstressed elastic solid, such as Hertz's, can therefore misattribute the force due to pretension to material stiffness.
Accounting for both is not straightforward either: analytical solutions for pre-stressed or pressurised sheets \cite{Bhatia1968IJNLM, Begley2004JMPS, Vella2017SM} show that the stiffness due to tension depends logarithmically on the size of the contact.
Adding the limiting laws for pretension and stretching can give large errors at intermediate depths~\cite{Vella2017SM}.
Second, the cortex is spatially heterogeneous, while AFM probes a finite region. 
The measured force may then depend on which part of the architecture lies beneath the probe, so that equal amounts of weakening give different responses and different architectures similar ones.

The above challenges motivate more mechanistically explicit models of the cortex. 
Continuum models describe the cortex as an active gel or viscous shell~\cite{joanny2013EPJE,kruse2005EPJE,prost2015nature,maxian2021plos}, and finite-element models have incorporated layered cell architectures, cortex thickness, substrate mechanics and indenter geometry~\cite{ferreira2020BMM,barreto2013biomat,krzemien2023BMM,VargasPinto2013BJ}.
These approaches capture cell-scale mechanics but average over the filament-scale structure through which local forces are transmitted.
Discrete network models resolve this structure and have established bending- and stretching-dominated regimes, nonlinear stiffening and a strong sensitivity to connectivity~\cite{mackintosh1995prl,head2003PRL,broedersz2014RMP,picu2011softmat,baumgarten2021softmat}.
Their relevance to indentation has been demonstrated by Merson \textit{et al.}, who showed that fibrous networks can deviate from Hertzian behaviour when the indenter size is not sufficiently larger than the mean network segment length~\cite{Merson2022JAM,Merson2023ActaBio}. 
In this regime, the measured response reflects mesh-scale deformation and local network rearrangement rather than an effective continuum modulus alone.

Existing approaches capture either the tension of the cortex or its network structure, but not both. 
AFM analyses that separate cortical tension from elasticity~\cite{Rivera2016BJ} and continuum solutions for pre-stressed and pressurised sheets~\cite{Bhatia1968IJNLM,Vella2017SM,Begley2004JMPS} treat the cortex as a homogeneous layer, whereas discrete network models of indentation~\cite{Merson2022JAM,Merson2023ActaBio} resolve the filament-scale structure but carry no pre-tension. 
How the spatial organisation of weakening within a tensioned, load-bearing network appears in an indentation measurement therefore remains open. 
The same amount of weakening may be spread throughout the network or concentrated into localised regions, and a local probe may rest on a weak region or on intact network. 
Whether these states can be told apart from indentation alone, and which features of the response carry information about them, is unresolved.

Here we address this with a microstructure-resolved model of a crosslinked, pressure-loaded cortex-like network indented by a micron-scale sphere, combined with a continuum membrane solution for the same geometry.
We ask why indentation stiffness is an ambiguous marker of the state of the cortex, in three parts. 
First, does a pre-tensioned fibre network respond to a probe like a tensioned membrane, so that its indentation force can be predicted from the tension of the layer alone? 
Second, at a fixed amount of weakening, how does the response depend on whether the weakening is diffuse or clustered, and on where it lies relative to the probe? 
Third, which features of the measurement, beyond a fitted stiffness, retain information about the architecture of the network?

The article is organized as follows: Section~\ref{sec:model} describes the discrete network model, and Section~\ref{sec:cont_descrip} the continuum description of the pressurized layer that serves as a reference.
Section~\ref{sec:results} presents the indentation results and their microscopic origin. 
Conclusions summarize the implications and limitations.

\section{Discrete network model}
\label{sec:model}
\subsection{Fibers and interactions}
The cell cortex is represented as a flat, non-woven, cross-linked fiber network representing a locally planar patch of cortical layer beneath the plasma membrane.
Individual fibers are modeled as coarse grained bundles of actin filaments, motivated by the tendency of cortical actin to form bundles through electrostatic and protein-mediated interactions~\cite{Nietmann2023Nature}. 
The network is treated as a mechanically discrete system whose response emerges from filament stretching, bending, excluded-volume interactions, and inter-fiber crosslinking.

%\subsubsection*{Coarse-grained fiber representation}
Each model fiber is taken to represent a bundle of approximately $n$ parallel actin filaments.
Assuming compact packing in the bundle cross-section, the total cross-sectional area of the fiber scales linearly with the bundling parameter $n$, hence the effective bundle diameter scales as: $d_{\text{f}}\approx d_{\text{a}}\sqrt{n/\phi_{\text{pack}}},$ where $\phi_{\text{pack}}\approx0.907$ is the packing efficiency for hexagonal close packing (HCP) and $d_{\text{a}}$, the diameter of a single actin filament, is around 7 nm~\cite{Nietmann2023Nature}.
In the coarse-grained representation, the mass of the actin material contained in a bundled filament segment of length $d_{\text{f}}$ is $m_{\text{b}}=n(d_{\text{f}}/2.76~\text{nm})m_{\text{actin}}$, where $m_{\text{actin}}\approx42~\text{kDa}$ is the mass of one actin monomer and 2.76 nm is the axial rise per subunit in F-actin. 

Each fiber is discretized into beads connected by harmonic bonds. 
Harmonic angle potentials between consecutive bonds provide its bending stiffness.
Crosslinks between adjacent fibers are represented by additional harmonic springs. 
The total bonded potential ${V}_{\text{b}}(r)$ can thus be expressed as:
\begin{equation}
    {V}_{\text{b}}(r)=\sum_i{\frac{k_{f}}{2}(r_{i}-r_{0,f})^2}+\sum_j{\frac{\kappa_{f}}{2}(\theta_j-\theta_{0,f})^2}+\sum_k{\frac{k_{\times}}{2}(r_k-r_{0,\times})^2},
\end{equation}
where $k_f$ is the axial spring stiffness of the fiber backbone, $\kappa_f$ is the angular stiffness associated with fiber bending, and $k_{\times}$ is the crosslink spring stiffness; $r_{0,f}$, $r_{0,\times}$, and $\theta_{0,f}$ denote the corresponding equilibrium bond lengths and equilibrium angle.
The axial stiffness of a single actin filament was estimated from direct measurements by Kojima et al.~\cite{kojima1994PNAS}, who reported a stiffness of 43.7–65.3 pN/nm for a 1 $\mu$m long filament. 
We use 58 pN/nm, between these two values.
Stiffness scales inversely with length. A bond of one bead spacing, $d_\text{f}=$50 nm, therefore has the stiffness $k_\text{actin}=$ (58 pN/nm)(1 $\mu$m/$d_\text{f}$) = 1160 pN/nm. For a coarse-grained fibre representing a bundle of n filaments loaded in parallel, the fibre stiffness is $k_\text{f} = n k_{\text{actin}}$.
%
%For a coarse-grained fiber representing a bundle of $n$ filaments loaded in parallel, the bundled fiber stiffness can be scaled as $k_f=nk_{\text{actin}}$, where $k_{\text{actin}}$($=(58~\text{pN/nm}\cdot 1\mu\text{m})/50~\text{nm} $) is the stiffness corresponding to the coarse-grained bead spacing.
%
The bending stiffness of actin is estimated using semiflexible polymer relation: $\kappa_{\text{actin}}=k_{\text{B}}Tl_{\text{p}}$, where $T$ is the temperature, $l_{\text{p}}$ the persistence-length, and $k_{\text{B}}$ the Boltzmann constant. 
For a bundled fiber, the effective bending rigidity is scaled as $\kappa_f=n^{\zeta}\kappa_{\text{actin}}$, where $\zeta$ accounts for the degree of inter-filament coupling within the bundle. 
In the weak-coupling limit, $\zeta=1$, whereas in strongly coupled limit, $\zeta=2$. 
We use $\zeta=1$ in the present study.

%\subsubsection*{Non-bonded interactions}
Excluded-volume interactions between beads are described using Lennard-Jones potential:
\begin{equation}
    V_{\text{nb}}(r) = 4\epsilon\left[\left(\frac{\sigma}{r}\right)^{12}-\left(\frac{\sigma}{r}\right)^{6}\right]
\end{equation}
The parameter $\sigma$ is found by setting equilibrium spacing as bead-spacing, $\sigma=2^{-1/6}d_{\text{f}}$. The parameter $\epsilon$ is found by setting the initial repulsive stiffness to that of $k_{\text{f}}$ as $\epsilon=k_{\text{f}} {d_{\text{f}}}^2/72$. The attractive tail has been cut-off to ensure only repulsion.

The set of actin-related parameters used in the simulations is summarized in Table~\ref{tab:sim_param}. All simulations are performed using a LAMMPS-based non-equilibrium molecular dynamics method.

\begin{table}[hbt!]
\caption{Fiber properties and other simulation parameters used in this study~\cite{Nietmann2023Nature,komori1977TRJ,kojima1994PNAS}. The simulations employ $\alpha$-actin filament properties. For fiber length, the reported mean and standard deviation correspond to a lognormal distribution. }
\label{tab:sim_param}
\centering
\begin{tabular}{c c l l}
\hline
\textbf{Parameter} & \textbf{Symbol} & \textbf{Value} & \textbf{Units} \\\hline
Fiber length~\cite{Nietmann2023Nature} &$l_{\text{f}}$ &$5.2\pm3.0$ & $\mu$m\\ %$\beta: 10.3\pm6.7$, $\gamma:12.8\pm10.9$ & $\mu$m \\
Bundling parameter~\cite{Nietmann2023Nature} &$n$ & 50 & \\
Bead diameter & $d_{\text{f}}$& 50 & nm \\
Actin persistence length~\cite{Nietmann2023Nature} &$l_{\text{p}}$&17 & $\mu$m \\
\hline 
Simulation box area &$A_{\text{box}}$&$50\times50$ & $\mu\text{m}^2$ \\
Unclamped network area &$A$&$40\times40$ & $\mu\text{m}^2$ \\
Target cortex thickness &$h$&240&$\text{nm}$\\
No. of fibers & $N_{\text{f}}$ & 6000 &\\
\hline
Bead mass~\cite{dominguez2011ARB,heissler2025FEBS} & $m_{\text{b}}$ &64& ag\\
Actin stiffness~\cite{kojima1994PNAS} & $k_{\text{actin}}$ &1160& pN/nm\\
Actin bending rigidity & $\kappa_{\text{actin}}$&439.5 & eV$\cdot$nm\\
\hline
\end{tabular}
\end{table}

\subsection{Network formation}
The number of fibers required to obtain the target network is estimated using the contact-density theory by Komori et. al.~\cite{komori1977TRJ}:
\begin{equation}
    N_{\text{f}} = \frac{\pi A_{\text{box}} h}{4d_{\text{f}}l_{\text{f}}\bar{l}},
\label{eq:fib_count}
\end{equation}
where $A_{\text{box}}$ is the area of network, while $h$ and $\bar{l}$ are the target cortex thickness and mean-segment-length (or mesh-size) respectively. Fiber lengths are assumed to follow a lognormal distribution~\cite{Nietmann2023Nature}.

A suspension of $N_{\text{f}}$ randomly oriented fibers is first generated in a 3-dimensional domain whose in-plane dimension equal the desired network size. 
The fibers are then deposited on a substrate using a moving substrate method~\cite{Agarwal2023MRC,Agarwal2024MSMSE}. Afterwards, the deposited fibers are pressed to the desired cortex thickness. 
During pressing, adjacent fibers are permitted to form crosslinks upon proximity, leading to an interconnected, mechanically stable network.

\subsection{Representation of cortical weakening}

We represent unhealthy cortical states by two defect architectures with the same overall degree of weakening: weak fibres distributed at random (diffuse weakening) and weak fibres gathered into clusters (localised weakening).
The first mimics a uniform loss of filament integrity, the second localised damage or remodelling. 
Both are plausible because the cortex is continually remodelled through actin turnover~\cite{fritzsche2016ScAdv} and its architecture regulates surface mechanics non-uniformly~\cite{chugh2017Nature}.

Weak fibres differ from healthy ones in stiffness and in rest shape. 
Their axial and bending stiffnesses are reduced to $0.2k_\text f$ and $0.2\kappa_\text f$, their rest bond length is increased from 50 to 60 nm, and their rest bond angle set to $160^\circ$ against $180^\circ$ in the healthy state. 
Since both axial and bending stiffness scale with the number of filaments in the bundle, the loss of stiffness corresponds to a bundle that has lost four-fifths of its filaments, for instance through severing.
The reduction in bending rigidity alone is also comparable to that caused by cofilin, which lowers the flexural rigidity of actin filaments about fivefold~\cite{McCullough2008JMB}. 
Geometry of the weak fibers corresponds to their waviness and excess contour-length. 
They thus represent slack, buckled filaments of the kind produced by myosin-driven compression in cortex-like networks, where buckled segments with radii of curvature of about 300 nm are prone to severing~\cite{Murrell2012PNAS}.
The rest angle used here corresponds to a radius of about 170 nm.

In the diffuse case, fibres are weakened at random until about 20\% of all intrafibre bonds are weak. 
In the clustered case, the same number of weak bonds is divided into $n_\text{clusters}$ clusters of $n_\text{wb}$ bonds each. 
Varying $n_\text{clusters}$ and $n_\text{wb}$ at constant product changes the cluster size without changing the total weakening (Table~\ref{tab:clust_weak}).
\begin{table}[hbt!]
\caption{The summary of number of weak clusters and weak bonds per clusters. The average defect diameter and their comparison with indenter diameter of 3 $\mu$m and mean-segment-length are also reported}
\label{tab:clust_weak}
\centering
\begin{tabular}{c c | c c c}
\hline
$n_{\text{cluster}}$& $n_{\text{wb}}$ & $\bar{d}_{\text{defect}}~(\mu$m) & $\bar{d}_{\text{defect}}/d_{\text{ind}}$ & $\bar{d}_{\text{defect}}/\bar{l}$  \\\hline
10                   & 10000                & 8.026                     & 2.675                                    & 30.4                              \\
15 & 6667 & 6.625 & 2.208 & 25.1 \\
20                   & 5000                 & 5.560                     & 1.853                                    & 21.1                              \\
40                   & 2500                 & 3.864                     & 1.288                                    & 14.6    \\
\hline
\end{tabular}
\end{table}

\subsection{Loading and boundary conditions}
The effects of intracellular pressure, myosin-induced cortical contractility, and mechanical coupling to the outer lipid membrane are incorporated in an effective manner through externally applied forces on the coarse-grained particles, which are described in subsequent paragraphs.

Intracellular pressure is modelled as a uniform out-of-plane pressure acting on the cortex-like network. 
Hydrostatic pressure drives mitotic cell rounding together with the actomyosin cortex~\cite{stewart2011nature}; in HeLa cells, the pressure excess increases from about 40 Pa in interphase to about 400 Pa in metaphase~\cite{fischer2014ScRep}. 
Guided by these values, we apply $p_\text{cell}=500$ Pa.

Myosin-II generates contractile cortical tension, which rises from about 0.2 mN/m in interphase to about 1.6 mN/m in metaphase~\cite{stewart2011nature,fischer2014ScRep}. 
For an approximately spherical cell of diameter $D$, the surface tension $\gamma$ is related to the effective pressure as $4\gamma/D$.
With $\gamma=1.6$ mN/m and $D\approx13$ µm, we represent myosin activity by an additional effective pressure $p_\text{myosin}=500$ Pa. 
This active contribution acts on top of the hydrostatic pressure, so the total load, $p=p_\text{cell}+p_\text{myosin}=1000$ Pa, represents a strongly tensioned mitotic rather than interphase cortex. 
The distinction matters, because the mechanical properties of the cortex change in opposite directions in interphase and mitosis upon oncogenic transformation~\cite{Hosseini2020AS}.

The out-of-plane displacement is fixed along the lateral boundaries, leaving a free square patch of side $L=40$ µm and area $A=L^2$ that is loaded by the pressure.
%
%Fig.~\ref{fig:network_pressure} shows the networks after loading.
%
%Healthy network in Fig.~\ref{fig:network_pressure}(a) inflates into a smooth dome. 
%
%The diffusely weakened network (Fig.~\ref{fig:network_pressure}(b)) keeps the dome shape but bulges further, its weak fibres (orange in the inset) wavier than the healthy ones 
%
%Clustered weak fibres with $\bar d_\text{defect}\approx1.853\,d_\text{ind}$, centred on or displaced from the indenter axis in Figs.~\ref{fig:network_pressure}(c,d), produce pronounced local protrusions and a rougher surface at the same pressure.
%

\subsection{Indentation and observables}
The network is indented by a rigid sphere of diameter $d_{\text{ind}}=3~\mu$m. 
A nano-sharp tip would probe the pore-scale structure of this sparse network and could pass through gaps between fibres, whereas a micron-scale sphere samples its mechanical response. 
Spheres with radii of 0.5--7~$\mu$m are also standard in cell indentation, where they avoid the overestimated moduli that sharp pyramidal tips give on soft samples~\citep{Sicard2017JMBBM,VargasPinto2013BJ}. 
The indentation nevertheless does not reach the continuum limit: with $d_{\text{ind}}/\bar l\approx8$ and 11 for the two mean segment lengths, it lies below the ratio $d_{\text{ind}}/\bar l\approx24$ above which Hertz-like behaviour has been reported for fibre networks without pre-tension~\citep{Merson2023ActaBio}. 
The probe therefore averages over individual fibres and pores while remaining sensitive to the network architecture, the regime in which spatially heterogeneous weakening can affect a local measurement.

We record the normal force $F$ with indentation depth $\delta$.
Unless stated otherwise, the normalized variables are $\tilde{F}=F/(pA)$ and $\tilde{\delta}=\delta/d_{\text{f}}$, where $pA=1600$~nN.
To compare the shapes of curves with different force levels, some defective responses are multiplied by a constant factor $\beta$ that brings them onto the healthy curve; this scaling is never used in fits.

The force--depth response is described in two ways. 
The first is the local exponent
\begin{equation}
\alpha=\frac{\mathrm{d}\ln\tilde F}{\mathrm{d}\ln\tilde\delta},
\label{eq:alpha_def}
\end{equation}
obtained by fitting $\tilde F\sim\tilde\delta^{\alpha}$ over moving windows of width $4d_{\text{f}}$ where the response is sufficiently smooth.
It assumes no functional form and is unchanged by the scaling with $\beta$. 
The second is a two-term representation. 
The force on a tensioned membrane contains a term linear in depth that is governed by the tension~\citep{Sen2005BJ}, and spherical contact gives a $\tilde\delta^{3/2}$ dependence; we therefore fit
\begin{equation}
\tilde{F}=a\tilde{\delta}+b\tilde{\delta}^{3/2}
\label{eq:twoterm}
\end{equation}
by least squares over $2\le\tilde\delta\le\tilde\delta_{\max}$, where $\tilde\delta_{\max}$ will be specified for each network while presenting the results.
Dividing by $\tilde\delta$ gives $\tilde F/\tilde\delta=a+b\sqrt{\tilde\delta}$, a straight line of intercept $a$ and slope $b$ on axes of $\tilde F/\tilde\delta$ against $\sqrt{\tilde\delta}$, which makes the adequacy of the form directly visible. 
Sums of limiting laws of this kind are common, but their accuracy depends on depth~\cite{Vella2017SM}.
We therefore use Eq.~\eqref{eq:twoterm} only to describe the response rather than as a decomposition into physical contributions, and interpret $a$ and $b$ by comparison with the membrane solution fitted over the same range.
Section~\ref{sec:healthy_macro} explains why the form nevertheless describes the present response well.
%
%Dividing by $\tilde\delta$ gives $\tilde F/\tilde\delta=a+b\sqrt{\tilde\delta}$, a straight line on axes of $\tilde F/\tilde\delta$ against $\sqrt{\tilde\delta}$, which makes the adequacy of the form directly visible. 
%

The local exponent of the two-term form is
\begin{equation}
\alpha=1+\frac{1}{2}\,\frac{b\sqrt{\tilde\delta}}{a+b\sqrt{\tilde\delta}},
\label{eq:alpha}
\end{equation}
which increases monotonically from 1 towards $3/2$ for positive $a$ and $b$.
Thus, a measured exponent that decreases with depth, with positive fitted $a$ and $b$, signals a departure from Eq.~\eqref{eq:twoterm}. 
Note that $a$ and $b$ are not independent and do not by themselves separate tension from elasticity, as will be shown in Section~\ref{sec:membrane_indentation}.

\section{Continuum description of pressurised layer}
\label{sec:cont_descrip}
This section treats the pressure-loaded network as a thin elastic sheet, first under inflation and then under indentation. 
Both problems are classical.
For the inflated sheet we use the Föppl–von Kármán equations~\cite{foppl1907,vonKarman1907} in Berger's approximation~\cite{berger1955JAM}, which replaces the varying in-plane stress by a single membrane tension, as summarized in Section~\ref{sec:methods_inflation}.
Sphere indentation of a pre-stressed membrane has been solved~\cite{Bhatia1968IJNLM, Begley2004JMPS}, and the pressurised membrane has been analysed for a contact of fixed radius~\cite{Vella2017SM}.
In Section~\ref{sec:membrane_indentation}, we derive the small-slope solution for a sphere pressed into a pressurised layer, where the contact radius grows with depth and the curvature of the inflated dome reduces the effective indenter radius.
The derivation gives closed-form expressions for the stiffness, its local exponent, the two-term coefficients and an inversion of the intercept for the in-plane modulus.
The only inputs to the indentation calculation are the membrane tension and n-plane modulus of the inflated layer, obtained from its profile before indentation (Section~\ref{sec:methods_inflation}). No indentation data enter, so the comparison with the simulations will be a test rather than a fit.
\subsection{Inflation and membrane tension}
\label{sec:methods_inflation}
Consider a sheet of thickness $h$, in-plane modulus $\bar{E}=E/(1-\nu^2)$ and bending rigidity $D=\bar{E}h^3/12$, held along the edges of a square of side $L$ and loaded by a uniform pressure $p$. 
For deflections small compared with $h$ the sheet bends, $D\nabla^4w=p$, and generates no in-plane tension~\cite{Timoshenko1959plates}; the corresponding three-dimensional linear-elastic solution of a slab is given in Appendix~\ref{app:continuum_sheet}. 
The ratio of stretching to bending energy, however, scales as $U_{\text{stretch}}/U_{\text{bend}}\sim12(1-\nu^2)(w/h)^2$, so stretching
dominates once the deflection exceeds the thickness.
Here $w_{\max}\simeq4~\mu$m and $h\approx240$~nm, so $w_{\max}/h\approx17$ and $U_{\text{stretch}}/U_{\text{bend}}\sim10^3$: the inflated layer is deep in the stretching-dominated regime.
We describe this stretching with the Berger approximation~\cite{berger1955JAM} to F\"oppl--von K\'arm\'an theory~\cite{foppl1907,vonKarman1907,Landau1986ToE}, in which the spatially varying in-plane
stress is replaced by a uniform, isotropic membrane force $N$ (force per unit length) set by the mean stretch of the sheet:
\begin{align}
D\nabla^4w-N\nabla^2w&=p,\label{eq:berger}\\
N&=\tfrac{1}{2}\bar{E}h\left\langle|\nabla w|^2\right\rangle,\label{eq:berger_N}
\end{align}
where $\langle\cdot\rangle$ denotes the average over the free sheet. 
Equation~\eqref{eq:berger_N} assumes that the in-plane displacement vanishes
along the edges, so that the mean stretch reduces to half the mean squared slope~\cite{berger1955JAM}; it thus also defines the mean stretch of the
inflated layer, $\varepsilon_0=N/(\bar{E}h)$.
For a prescribed $N$, Eq.~\eqref{eq:berger} is linear; we solve the pair self-consistently by finite differences, as detailed in Appendix~\ref{app:continuum}.

Away from a boundary layer of width $\ell_c=\sqrt{D/N}$, the bending term in Eq.~\eqref{eq:berger} is negligible and the sheet acts as a membrane. 
To obtain a closed-form relation between tension and modulus, we approximate the patch by a circular membrane of radius $R_{\text{m}}=L/2$. 
Vertical force balance on a disc of radius $r$ concentric with the patch, loaded by the pressure and by an axial point load $F$ that will represent the indenter, reads
\begin{equation}
2\pi rN\frac{\mathrm{d}W}{\mathrm{d}r}+\pi r^2p-F=0,
\label{eq:disc_balance}
\end{equation}
where $W(r)$ is the height of the layer. For the inflated state, $F=0$, so that $\mathrm{d}W_0/\mathrm{d}r=-pr/(2N)$ and
\begin{equation}
W_0(r)=\frac{p\,(R_{\text{m}}^2-r^2)}{4N}.
\label{eq:dome}
\end{equation}
The mean squared slope of this dome over the disc is
\begin{equation}
\left\langle|\nabla W_0|^2\right\rangle=\left(\frac{p}{2N}\right)^2\langle r^2\rangle=\left(\frac{p}{2N}\right)^2\frac{R_{\text{m}}^2}{2}=\frac{p^2R_{\text{m}}^2}{8N^2},
\end{equation}
and inserting it into Eq.~\eqref{eq:berger_N} gives $N=\bar{E}h\,p^2R_{\text{m}}^2/(16N^2)$, that is, with $R_{\text{m}}=L/2$,
\begin{equation}
N=\left(\frac{\bar{E}h\,p^2R_{\text{m}}^2}{16}\right)^{1/3}=\frac{1}{4}(\bar{E}h)^{1/3}(pL)^{2/3}.
\label{eq:N_analytic}
\end{equation}
The tension depends on the cube root of the in-plane modulus: a softer layer bulges further, and the larger slope largely restores its tension. 
This weak dependence will turn out to control how strongly global softening is felt by an indenter.

\subsection{Indentation of the pressurised membrane}
\label{sec:membrane_indentation}
\begin{figure}[hbt!]
    \centering
    \includegraphics[width=0.98\linewidth]{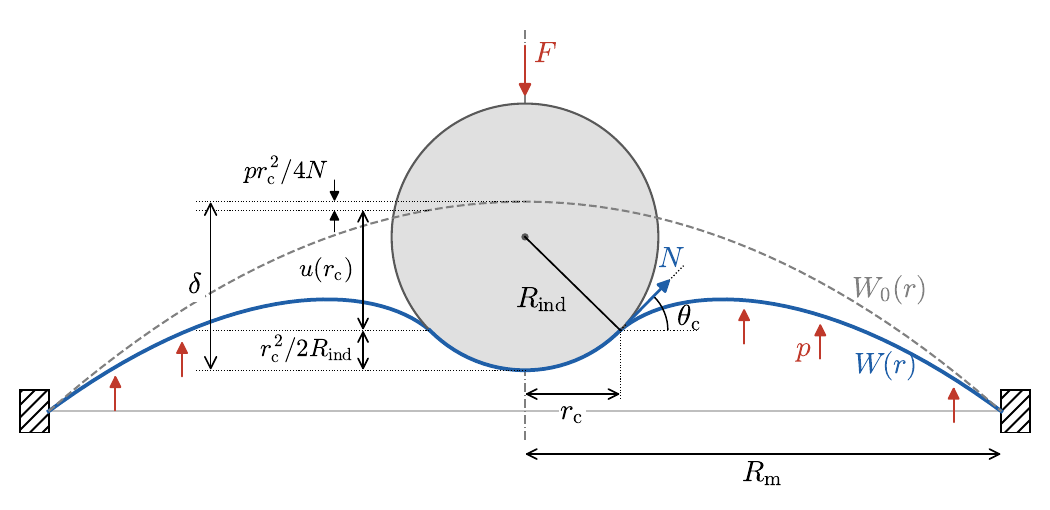}
    \caption{Schematic of the geometry of the membrane model. The pressure $p$ inflates the clamped layer of radius $R_\mathrm{m}$ into the dome $W_0(r)$ (dashed); a sphere of radius $R_\mathrm{ind}$ pressed with force $F$ produces the indented profile $W(r)$ (solid). The layer leaves the sphere tangentially at the contact radius $r_\mathrm{c}$, at angle $\theta_c$ with $\sin\theta_c = r_\mathrm{c}/R_\mathrm{ind}$, where the membrane force $N$ balances $F$. The indentation depth $\delta$ is sum of the fall of the dome $pr_c^2/(4N)$, the deflection $u(r_\mathrm{c})$ at the contact edge, and the cap height $r_\mathrm{c}^2/2R_\mathrm{ind}$.}
    \label{fig:membrane_geometry}
\end{figure}

A rigid sphere of radius $R_{\text{ind}}=d_{\text{ind}}/2$ is now pressed on the axis of the inflated membrane with force $F$ (Fig.~\ref{fig:membrane_geometry}). 
We keep $N$ at its inflated value and assume small slopes. %; both approximations are assessed in Section~\ref{sec:accuracy}. 
At constant $N$, Eq.~\eqref{eq:disc_balance} is linear, so the pressure continues to be carried by the dome $W_0$ and the indentation adds a deflection $u$, with $W=W_0-u$ outside the contact radius $r_{\text{c}}$. 
Subtracting the balance for $F=0$ from Eq.~\eqref{eq:disc_balance} leaves $2\pi rN\,\mathrm{d}u/\mathrm{d}r=-F$, and integrating with $u(R_{\text{m}})=0$ gives the logarithmic dimple
\begin{equation}
u(r)=\frac{F}{2\pi N}\ln\frac{R_{\text{m}}}{r},\qquad r\ge r_{\text{c}}.
\label{eq:dimple}
\end{equation}

At the contact edge the layer leaves the sphere tangentially, so its slope equals that of the sphere, $\mathrm{d}W/\mathrm{d}r=r_{\text{c}}/R_{\text{ind}}$. 
With $\mathrm{d}W/\mathrm{d}r=\mathrm{d}W_0/\mathrm{d}r-\mathrm{d}u/\mathrm{d}r=-pr_{\text{c}}/(2N)+F/(2\pi Nr_{\text{c}})$, this condition reads
\begin{equation}
\frac{F}{2\pi Nr_{\text{c}}}-\frac{pr_{\text{c}}}{2N}=\frac{r_{\text{c}}}{R_{\text{ind}}},
\end{equation}
which gives the contact force
\begin{equation}
F=\frac{2\pi Nr_{\text{c}}^2}{R^*},
\qquad\text{with}\qquad 
\frac{1}{R^*}=\frac{1}{R_{\text{ind}}}+\frac{p}{2N}.
\label{eq:contact_force}
\end{equation}
The effective radius $R^*$ accounts for the pressure acting on the contact disc: the curvature of the inflated dome, $p/2N$, adds to that of the sphere.
Measured from the apex of the dome, the indentation depth $\delta$ has three components: the fall of the dome itself, $pr_c^2/(4N)$ between the axis and contact edge; the displacement of the contact edge, $u(r_c)$; and the cap height below the contact edge, $r_c^2/(2R_{\text{ind}})$.
Using Eq.~\eqref{eq:contact_force} for $F$ in Eq.~\eqref{eq:dimple},
\begin{equation}
\delta=\frac{r_{\text{c}}^2}{R^*}\,\chi,
\qquad\text{with}\qquad 
\chi=\ln\frac{R_{\text{m}}}{r_{\text{c}}}+\frac{1}{2}.
\label{eq:depth}
\end{equation}
Dividing Eq.~\eqref{eq:contact_force} by Eq.~\eqref{eq:depth} gives the membrane stiffness
\begin{equation}
\frac{F}{\delta}=\frac{2\pi N}{\chi}.
\label{eq:f_delta}
\end{equation}
The depth enters through $\chi$. 
Eliminating $r_{\text{c}}^2=R^*\delta/\chi$ from the definition of $\chi$ gives 
\begin{equation}
2\chi-1-\ln\chi=\ln\frac{R_{\text{m}}^2}{R^*\delta}.
\label{eq:chi}
\end{equation}
The left-hand side has a minimum at $\chi=1/2$; since $r_c<R_m$, only the root with $\chi>1/2$ is physical.
On this branch $\chi$ decreases monotonically with depth.
The local exponent of the membrane response follows as
\begin{equation}
\alpha_{\text{N}}=\frac{\mathrm{d}\ln F}{\mathrm{d}\ln\delta}=1+\frac{1}{2\chi-1}=\frac{2\chi}{2\chi-1}.
\label{eq:alpha_m}
\end{equation}
Since $\chi$ decreases slowly with depth, a membrane at constant tension produces a force that is nearly linear in depth, with an exponent that increases only weakly, from 1.12 at $\tilde{\delta}=$5 to 1.14 at $\tilde{\delta}=$15 for the present geometry.

Since  $\chi$ varies logarithmically with depth, the membrane response is not exactly of the two-term form (Eq.~\eqref{eq:twoterm}). 
Its local intercept and slope nevertheless define the two-term coefficients of the membrane. 
With $\tilde F/\tilde\delta=(2\pi Nd_{\text{f}}/pA)/\chi$ and $\mathrm{d}\chi/\mathrm{d}\tilde\delta=-\chi/[\tilde\delta(2\chi-1)]$ from Eq.~\eqref{eq:chi}, the slope on these axes is
\begin{equation}
\frac{\mathrm{d}(\tilde F/\tilde\delta)}{\mathrm{d}\sqrt{\tilde\delta}}=2\sqrt{\tilde\delta}\,\frac{\mathrm{d}(\tilde F/\tilde\delta)}{\mathrm{d}\tilde\delta}=\frac{2\pi Nd_{\text{f}}}{pA}\,\frac{2}{\chi(2\chi-1)\sqrt{\tilde\delta}}.
\label{eq:slope_N}
\end{equation}
A straight line touching the membrane curve at a depth $\delta_0$ has this slope and the intercept $\tilde F/\tilde\delta-b_{\text{N}}\sqrt{\tilde\delta_0}$, so that
\begin{equation}
a_{\text{N}}=\frac{2\pi Nd_{\text{f}}}{pA}\,\frac{2\chi_0-3}{\chi_0(2\chi_0-1)},\qquad
b_{\text{N}}=\frac{2\pi Nd_{\text{f}}}{pA}\,\frac{2}{\chi_0(2\chi_0-1)\sqrt{\tilde\delta_0}},
\label{eq:ab_membrane}
\end{equation}
with $\chi_0=\chi(\delta_0)$. Both coefficients are proportional to $N$: the tension of the layer contributes not only the linear term but also part of the $\tilde\delta^{3/2}$ term, which is therefore not purely elastic.
Since the two coefficients are depth dependent, we fit Eq.~\eqref{eq:twoterm} to the membrane solution by least squares over the same depth range as each simulation.

The above solution assumes a constant membrane tension over the whole patch.
However, the dimple of Eq.~\eqref{eq:dimple} is steep near the contact. 
Its slope, $\mathrm{d}u/\mathrm{d}r=-F/(2\pi Nr)=-r_{\text{c}}^2/(R^*r)$, stretches the layer locally by
\begin{equation}
\varepsilon_{\text{ind}}(r)=\frac{1}{2}\left(\frac{\mathrm{d}u}{\mathrm{d}r}\right)^2=\frac{1}{2}\left(\frac{r_{\text{c}}^2}{R^*r}\right)^2=\varepsilon_0\left(\frac{r^{\ast}}{r} \right)^2,
\label{eq:eps_ind}
\end{equation}
where $r^{\ast}=r_{\text{c}}^2/(R^*\sqrt{2\varepsilon_0})$, is the radius within which indentation at least doubles the stretch of the layer.
To estimate the resulting stiffening, we let the local tension follow the local stretch, $N_{\text{loc}}=\bar Eh(\varepsilon_0+\varepsilon_{\text{ind}})=N[1+(r^*/r)^2]$, and use it in the force balance outside the contact, $2\pi rN_{\text{loc}}|\mathrm{d}u/\mathrm{d}r|=F$. 
Integrating $|\mathrm{d}u/\mathrm{d}r|=(F/2\pi N)\,r/(r^2+r^{*2})$ from $r_{\text{c}}$ to $R_{\text{m}}\gg r^*$ gives a shallower dimple, and the stiffness becomes
\begin{equation}
\frac{F}{\delta}=\frac{2\pi N}{\chi-\Delta\chi},\qquad
\Delta\chi=\frac{1}{2}\ln\left(1+\frac{r^{*2}}{r_{\text{c}}^2}\right),\qquad
\frac{r^{*2}}{r_{\text{c}}^2}=\frac{\delta}{2\varepsilon_0\chi R^*}.
\label{eq:dchi}
\end{equation}
Since $(r^{*}/r_{\text{c}})^2$ grows almost linearly with depth, the correction is negligible at small depths and grows with depth, so that it adds mainly to $b$. 
Its two-term coefficients, $a_\mathrm{m}$ and $b_\mathrm{m}$, are obtained in the same way as for the constant-tension solution, by fitting Eq.~\eqref{eq:twoterm} to Eq.~\eqref{eq:dchi} over the same depth range as the simulation.
The consequence of this correction over constant tension assumption will be discussed later in the results section.
The estimate computes the extra stretch from the slope of uniform tension solution, although the correction itself makes the dimple shallower, and it neglects the in-plane relaxation, which would reduce the local stretch.
It therefore identifies the mechanism and the form of the response rather than the exact magnitude.

%%%%%%%%%
\subsection{Inversion for the in-plane modulus}
\label{sec:modulus_invert}

Since $a_{\text{N}}$ is proportional to $N$, and $N$ is related to $\bar E$ by Eq.~\eqref{eq:N_analytic}, a measured intercept $a$ can be inverted for the in-plane modulus. 
With $A=L^2$, inverting Eq.~\eqref{eq:ab_membrane} and equating $a_\text N$ to the measured $a$ gives $N=a_{\text{N}}\,pL^2/(4d_{\text{f}}G_0)$.
Using Eq.~\eqref{eq:N_analytic}, we can now find
\begin{equation}
\bar{E}_a=\frac{pL}{h}\left(\frac{a}{G_0}\frac{L}{d_{\text{f}}}\right)^3,
\qquad \text{with} \qquad
G_0=\frac{\pi}{2}\,\frac{2\chi_0-3}{\chi_0(2\chi_0-1)}.
\label{eq:Ebar_from_a}
\end{equation}
Here the subscript indicates that the modulus is obtained from the inversion from the intercept $a$.
We average Eq.~\eqref{eq:Ebar_from_a} over $4\le\delta_0/d_{\text{f}}\le16$; applied to the membrane's own $a_{\text{N}}$, this recovers von K\'arm\'an modulus to within 3-9\%.
Since $E_a\propto a^3$, relative errors are tripled.
Only the intercept is inverted. 
Within the membrane model, the two coefficients are not independent: $b_{\text{N}}=2\,a_{\text{N}}/((2\chi_0-3)\sqrt{\tilde\delta_0})$.
Inverting $b_{\text{N}}$ would therefore add no information about $N$, only additional sensitivity to $\delta_0$.
Moreover, the linear term describes the response of the layer at its existing tension, whereas the measured $b$ also collects stiffening that grows with depth, such as the local stretching of the layer near the contact; inverting $b$ may therefore give a biased modulus.
The inversion uses the constant tension solution rather than the local stretching estimate due to its simplicity and controlled approximations.
%; the tho solutions lie within 1-2\% for $\tilde{\delta}<2$.
%

\section{Results}
\label{sec:results}
\subsection{Pressure-loaded networks as a mechanical reference}
\label{sec:results_inflate}
\begin{figure*}[hbt!]
    \centering
    \includegraphics[width=0.9\linewidth]{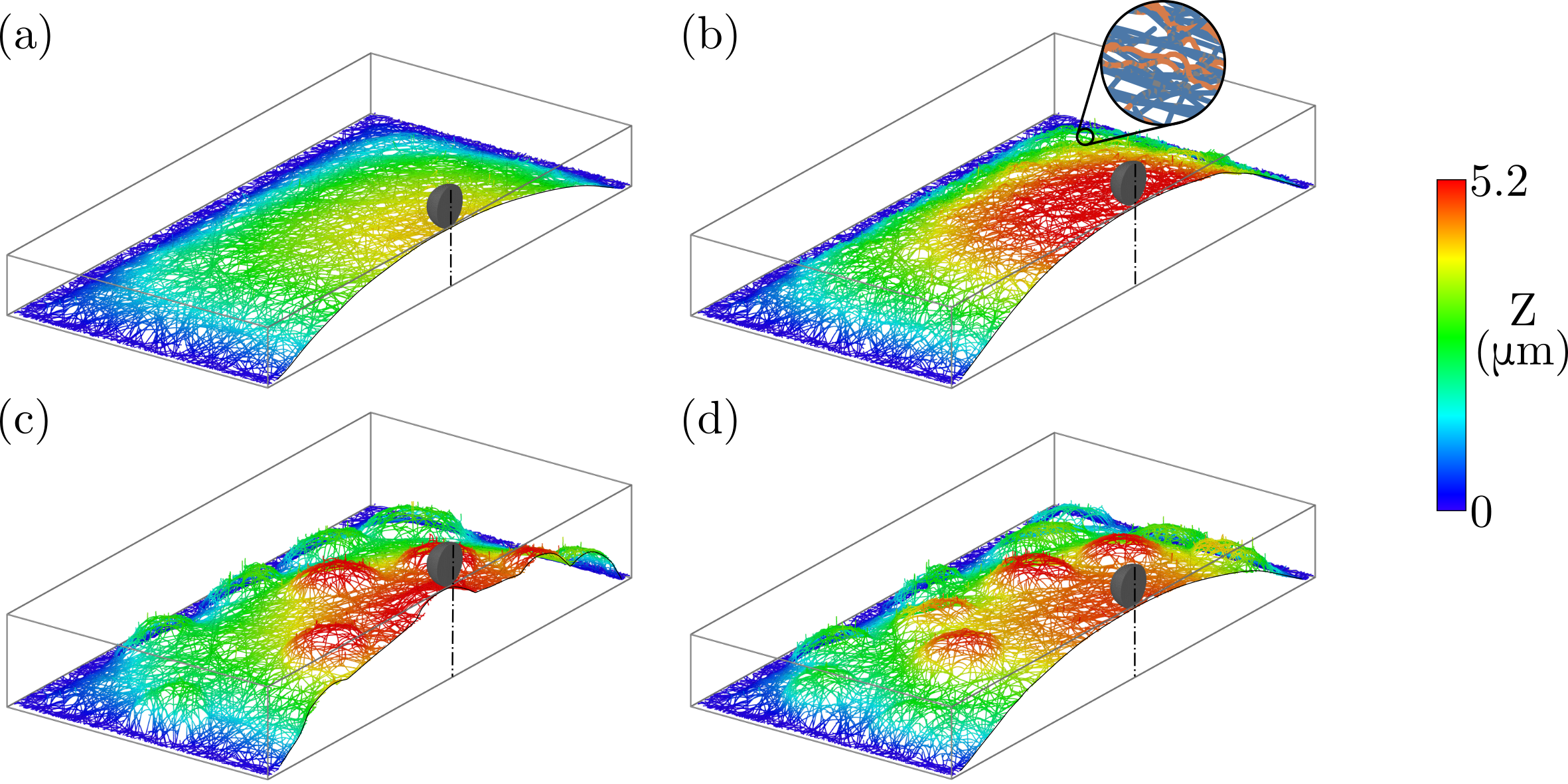}
    \caption{Pressure-loaded topography of healthy and defective cortex networks, before indentation, colored by out-of-plane displacement for (a) Healthy network; (b) Network with diffusely distributed weak fibers, (c) Networks with centered-clustered weak fibers, (d) Networks with off-centered-clustered weak fibers. The clusters in (c,d) have diameter of $\bar{d}_{\text{defect}}\approx1.853d_{\text{ind}}$.}
    \label{fig:network_pressure}
\end{figure*}

Fig.~\ref{fig:network_pressure} compares the pressure loaded reference profile of healthy and defective networks before indentation.
Healthy network in Fig.~\ref{fig:network_pressure}(a) inflates into a smooth dome.
The diffusely weakened network keeps the dome shape but bulges further, and its weak fibres are visibly more wavy (Fig.~\ref{fig:network_pressure}b).
In contrast, the clustered weakening, centred on or displaced from the indenter axis in Figs.~\ref{fig:network_pressure}(c,d) produces pronounced protrusions and spatial variation in height.

\begin{figure}[hbt!]
    \centering
    \includegraphics[width=0.95\linewidth]{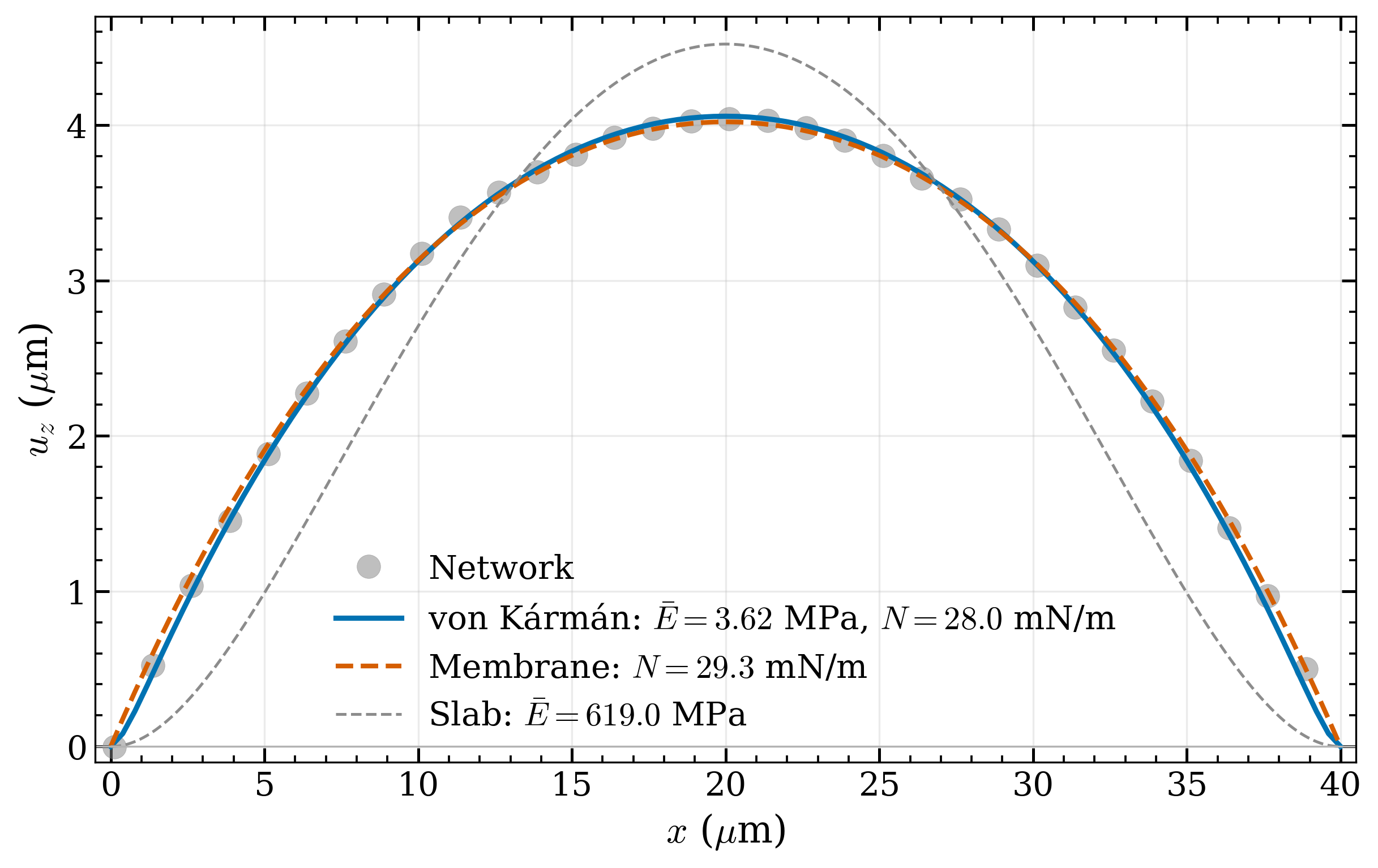}
    \caption{Out-of-plane deformation of the healthy network under pressure, compared with continuum solutions. Symbols: coarse-grained network profile. Solid line: F\"oppl-von K\'arm\'an solution in Berger's approximation, with  $\bar{E}=E/(1-\nu^2)$ fitted. Orange dashed line: membrane with tension only, with $N$ fitted. Grey dashed line: linear elastic slab, with $\bar E$ fitted. The membrane and the von Kármán solution coincide, except within $\ell_\mathrm{c}$ of the clamped edges. The linear slab has the wrong shape and needs a modulus about 170 times larger.}
    \label{fig:vK_fit}
\end{figure}

The Föppl–von Kármán solution in Berger’s approximation (Section~\ref{sec:methods_inflation}) reproduces the coarse-grained profile of the healthy network closely with $\bar{E}=3.62$~MPa (Fig.~\ref{fig:vK_fit}).
In contrast, the linear elastic-slab solution has a markedly different shape and a high fitted modulus of $\bar{E}=619$~MPa.
The difference is geometric. The peak deflection, $w_\mathrm{max} \approx 4.0~\mu$m, is about 17 times the layer thickness. 
Stretching, not bending, carries the load.
The fitted membrane force, $N=28.0$~mN/m, is consistent with Laplace's law, $N=pR/2$, which with the radius of curvature of the dome, $R\approx L^2/(8w_{\max})\approx50~\mu$m, gives 25~mN/m. 
A membrane with tension alone fits the profile equally well, with $N = 29.3$ mN/m (Fig.~\ref{fig:vK_fit}).
Bending matters only within $\ell_c=\sqrt{D/N}\approx0.39~\mu$m of the edges, about 1\% of the span, so the profile is insensitive to whether the edges are clamped or simply supported. 
Although it is microscopically discrete, the network therefore behaves globally as a strongly stretched membrane.

The profile alone fixes the tension but not the modulus. 
The von Kármán description adds the link between them: the tension is generated by stretching (Eq.~\eqref{eq:berger_N}).
This link gives $\bar E$, which we need for local stretching during indentation, as will be shown in the next subsection.
In a denser, stiffer network, bending would become visible in the profile, because $\ell_\mathrm{c} \propto \bar E^{1/3}$ at fixed pressure. 
Here, $\ell_\mathrm{c}$ would reach 10\% of the span only for an $\bar E$ about 1000 times larger. 
This would require about a thousand times more fibres, which is beyond the reach of the present simulations.

With $\bar{E}=3.62$~MPa, the circular-membrane relation, Eq.~\eqref{eq:N_analytic}, gives $N=27.9$~mN/m, against 28.0~mN/m for the square patch, so the circular approximation retains the tension of the layer. 
The same fit applied to each defective network gives in-plane moduli between 1.2 and 1.8~MPa (Table~\ref{tab:membrane}). 
%
%These tensions, taken from the inflated profiles alone, are the only input to the membrane predictions below.

\subsection{Healthy-network indentation}
\label{sec:healthy_macro}
\begin{figure}[hbt!]
    \centering
    \includegraphics[width=0.95\linewidth]{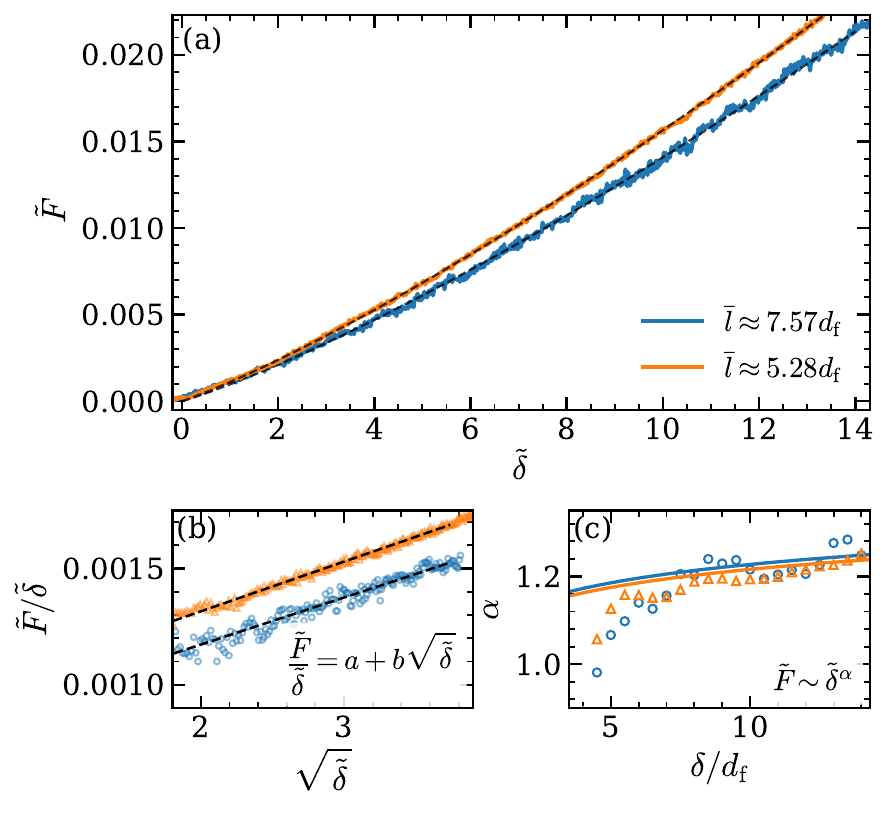}
    \caption{Indentation response of healthy networks at two mean segment lengths. (a) Normalised force $\tilde F = F/(pA)$ against normalised depth $\tilde\delta = \delta/d_{\rm f}$, where $pA = 1600$~nN is the force due to the applied pressure and $d_{\rm f}$ the bead diameter; the indenter diameter is $3~\mu$m. Dashed lines are fits of the two-term form $\tilde F = a\tilde\delta + b\tilde\delta^{3/2}$, giving $a = 7.68\times10^{-4}$, $b = 2.02\times10^{-4}$ for $\bar l \approx 7.57 d_{\rm f}$ and $a = 8.89\times10^{-4}$, $b = 2.14\times10^{-4}$ for $\bar l \approx 5.28 d_{\rm f}$. (b) The same data as $\tilde F/\tilde\delta$ against $\sqrt{\tilde\delta}$, on which the two-term form is a straight line of intercept $a$ and slope $b$; dashed lines are the fits of panel (a). (c) Local exponent $\alpha$ obtained by fitting $\tilde F \sim \tilde\delta^{\alpha}$ over moving windows of width $4 d_{\rm f}$ (symbols), with the prediction $\alpha = 1 + \tfrac{1}{2} b\sqrt{\tilde\delta}/(a + b\sqrt{\tilde\delta})$ of the fitted coefficients (lines). Colours are as in panel (a).}
    \label{fig:indent_healthy}
\end{figure}

Fig.~\ref{fig:indent_healthy}(a) shows the simulated-indentation response of the healthy network for two mean segment length (mesh-sizes).
Both responses exhibit monotonic increase in normalized indentation force with normalized depth.
The forces remain close at small depths; beyond $\delta\approx2d_{\text{f}}$ the more densely crosslinked network carries a higher force.
Also the sparser crosslinked network responds intermittently, presumably through local rearrangements.

On axes of $\tilde F/\tilde\delta$ against $\sqrt{\tilde\delta}$, both networks follow a straight line over the whole depth range (Fig.~\ref{fig:indent_healthy}b), so the two-term form well describes the healthy response.
The respective fits for the two term coefficients by least squares give $(a,b)=(7.68,2.02)\times10^{-4}$ for the sparsely-connected and $(8.89,2.14)\times10^{-4}$ for the densely-connected networks.
Thus the latter exhibits a larger linear coefficient (by $\approx16\%$) while the nonlinear coefficient changes only modestly (by $\approx6\%$).
The local exponent $\alpha$ increases continuously with depth (Fig.~\ref{fig:indent_healthy}c) for simulation data and follows the one predicted by two-term (Eq.~\eqref{eq:alpha}) for $\tilde\delta>7$.
Thus, no single power law describes the response.

\begin{figure}[hbt!]
    \centering
    \includegraphics[width=0.98\linewidth]{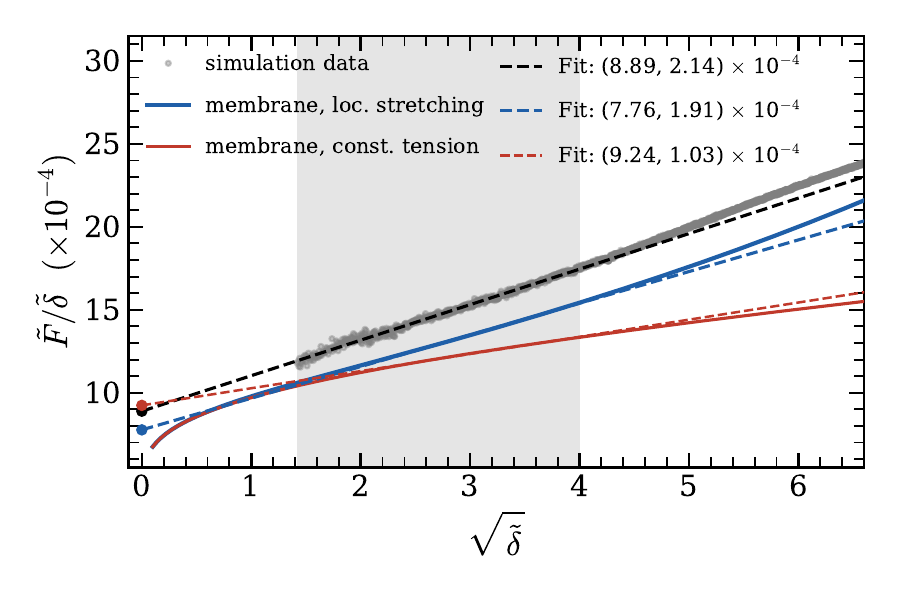}
    \caption{Comparison of simulation results for healthy network (gray symbols) with membrane-model predictions incorporating local stretching (blue) and constant tension (red). Solid lines represent analytical solutions, while dashed lines denote the corresponding two-term fits (Eq.~\eqref{eq:twoterm}), including the fit to simulation data (black). The fitted parameters (a,b) are given in parentheses in the legend. The shaded region indicates the fitting range.}
    \label{fig:healthy_twoterm_membrane}
\end{figure}

Fig.~\ref{fig:healthy_twoterm_membrane} compares the simulated response of the denser network with two membrane solutions, both evaluated with the tension of the inflated layer, $N=28.0$ mN/m and $\bar{E}=3.62$~MPa, with no other adjustable parameter. 
One keeps this tension uniform (red), while the other lets it rise near the contact (blue).
At constant tension (red; Eqs.~\eqref{eq:f_delta} and~\eqref{eq:chi}), the response is slightly concave on these axes because $\chi$ varies logarithmically with depth. 
Its two-term fit over the shaded range, $(a_\mathrm N,b_\mathrm N)=(9.24,1.03)\times10^{-4}$, matches the simulated intercept $a=8.89\times10^{-4}$ to within 4\%, but only half of the slope $b=2.14\times10^{-4}$.
The difference $b-b_{\text{N}}=1.11\times10^{-4}$, measures how much more the network stiffens with depth than the membrane at constant tension; we call it excess-stiffening.
Allowing the tension to rise near the contact (blue, Eq.~\eqref{eq:dchi}) closes about 75\% of this gap, $(a_\mathrm{m},b_\mathrm{m})=(7.76,1.91)\times10^{-4}$. 
The corrected curve lies 10–12\% below the simulation throughout, but reproduces the form of the response, including its upward curvature beyond the fitted range, which the constant-tension solution lacks. 
For $\tilde\delta\lesssim20$, which covers the depths analyzed here, both the simulation and analytical solution follow the two-term form reasonably well.
Throughout, we use $b-b_{\text{N}}$ rather than $b-b_{\text{m}}$ as a measure of excess-stiffening.
Since $b_\text N$ is exact for a membrane at uniform tension, this measure does not depend on the approximations of the local stretching estimate, which we use only to explain the excess.

Why do the simulation and corrected solution follow the two-term form so closely, although neither contains a Hertzian term?
Differentiating Eq.~\eqref{eq:dchi} with respect to $\sqrt{\tilde{\delta}}$, using Eq.~\eqref{eq:chi} gives
\begin{equation}
    \frac{\mathrm d(\tilde F/\tilde\delta)}{\mathrm d\sqrt{\tilde\delta}}=\frac{2\pi Nd_\mathrm f}{pA}\,\frac{\alpha_\mathrm N}{\sqrt{\tilde\delta}\,(\chi-\Delta\chi)^2}\left[1+\frac{r^{*2}}{r_\mathrm c^2+r^{*2}}\right],
    \label{eq:slope_LS}
\end{equation}
which reduces to Eq.~\eqref{eq:slope_N} at constant tension ($\Delta \chi=0,r^{\ast}=0$).
Eq.~\eqref{eq:slope_LS} is a product of factors that change with depth in opposite directions.
The factor $1/\sqrt{\tilde{\delta}}$ decreases with depth: it dominates at constant tension, so the slope of the red curve in Fig.~\ref{fig:healthy_twoterm_membrane} falls.
The factor $(\chi-\Delta \chi)^{-2}$ increases as the contact grows and the bracketed term rises from 1 to 2 with the depth.
Over the simulated depths the decrease and the increase nearly cancel: between $\tilde\delta=1$ and 14, $1/\sqrt{\tilde\delta}$ falls by a factor of 3.7, while the other factors together rise by a factor of 3.6.
The response is therefore close to a straight line.
At larger depths the increasing factors win, and the response bends upward, as both the simulation and the corrected solution do beyond the fitted range in Fig.~\ref{fig:healthy_twoterm_membrane}.
The local stretching solution also reproduces the depth dependence of measured local exponent $\alpha$.
For the healthy network it gives $\alpha=$1.17,1.22, and 1.25 at $\tilde{\delta}=$5, 10, and 14, against 1.14, 1.21, and 1.26 from the simulation, whereas for the constant-tension solution it stays between 1.12 and 1.14.

\subsubsection{Diffuse weakening}
\label{sec:diffuse_macro}
\begin{figure}[hbt!]
    \centering
    \includegraphics[width=0.95\linewidth]{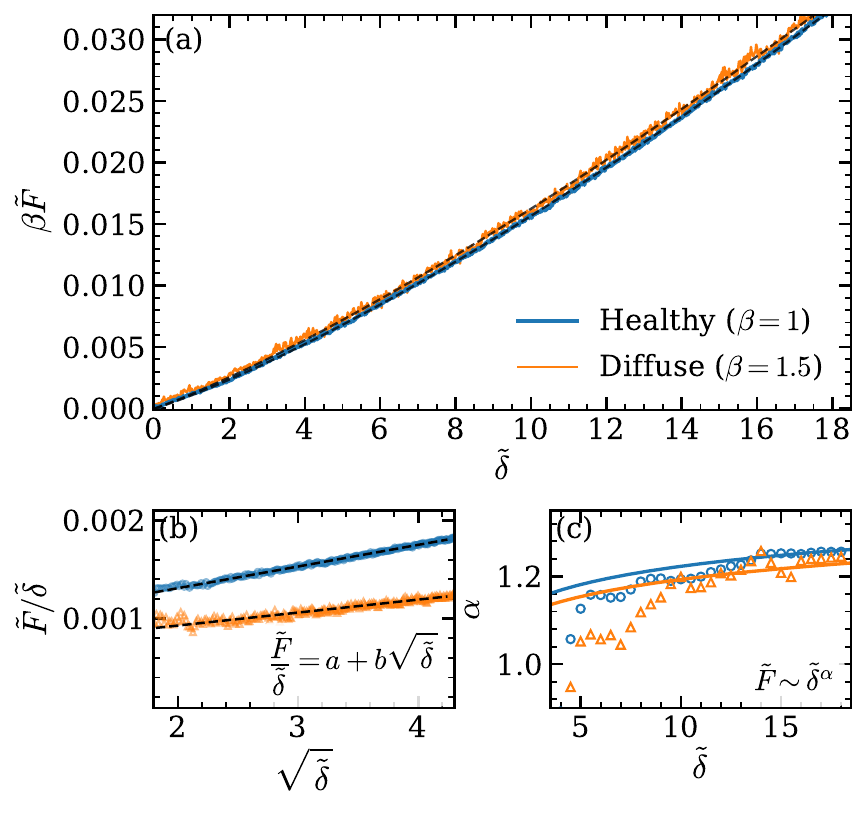}
    \caption{Indentation response of the diffusely weakened network compared with
the healthy reference. (a) Normalised force $\tilde F = F/(pA)$ against
normalised depth $\tilde\delta = \delta/d_{\rm f}$, with the defective response
scaled by $\beta = 1.5$ to bring it onto the healthy curve. Dashed lines are
fits of $\tilde F = a\tilde\delta + b\tilde\delta^{3/2}$. The fits give $a = 8.89\times10^{-4}$, $b = 2.14\times10^{-4}$ for the
healthy network and $a = 6.65\times10^{-4}$, $b = 1.32\times10^{-4}$ for the
weakened one, corresponding to crossover depths
$\tilde\delta^{*} = (a/b)^{2}$ of $17.3$ and $25.3$. (b) The unscaled data as
$\tilde F/\tilde\delta$ against $\sqrt{\tilde\delta}$, on which the two-term
form is a straight line of intercept $a$ and slope $b$; the vertical offset
between the two sets reflects the difference in $a$ and the difference in slope
that in $b$. (c) Local exponent $\alpha$ from fits of
$\tilde F \sim \tilde\delta^{\alpha}$ over moving windows of width
$4 d_{\rm f}$ (symbols), with the prediction
$\alpha = 1 + \tfrac{1}{2} b\sqrt{\tilde\delta}/(a + b\sqrt{\tilde\delta})$ of
the fitted coefficients (lines). Colours as in panel (a).}
    \label{fig:indent_weak_diffused}
\end{figure}

We first consider the case of diffusely distributed weak fibers, a reference pathological state corresponding to spatially uniform cortical weakening.
Fig.~\ref{fig:indent_weak_diffused}(a) compares the indentation response of this defective network with the healthy one. 
Multiplying the force of weak network by $\beta=1.5$ approximately brings it to the same force scale as the healthy network. 
Its response follows a straight line on axes of $\tilde F/\tilde\delta$ against $\sqrt{\tilde\delta}$ (Fig.~\ref{fig:indent_weak_diffused}(b)), with $(a,b)=(6.65,1.32)\times10^{-4}$.
The lower force reflects the lower tension of the weakened layer.
Its inflated profile gives an in-plane modulus 2.5 times lower than that of the healthy network, 1.42 against 3.62~MPa.
Since the tension scales with the cube root of the modulus, Eq.~\eqref{eq:N_analytic}, $N$ is only 26\% lower. 
The intercept follows the same scaling: it is 1.34 times lower than in healthy network against $\sqrt[3]{2.55}\approx 1.37$ predicted from the moduli.
With this tension, the membrane predicts $a_{\text{N}}=6.86\times10^{-4}$, 25\% below the healthy value and within 3\% of the simulated intercept.
Inverting the intercept recovers the modulus, $\bar E_a=1.37$ against 1.42 MPa (Table~\ref{tab:membrane}).
The excess stiffening is also smaller, $b-b_{\text{N}}=0.57\times10^{-4}$ against $1.07\times10^{-4}$ for the healthy network, consistent with the weaker increase in bond tension beneath the probe. 

The local exponent $\alpha$ in Fig.~\ref{fig:indent_weak_diffused}(c) follows a similar depth dependence and positive slope for the two networks.
The weakened one shows slightly lower values than healthy one initially, consistent with its lower $b/a$ value.
But the two almost converge for $\tilde{\delta}>10$. 
Therefore, the local exponent distinguishes the two networks only weakly.

\subsubsection{Clustered weakening: centered defect}
\label{sec:cluster_c}

We now consider the networks in which weak fibers are spatially clustered rather than diffusely distributed. 
In all clustered cases, the total number of weak bonds is kept fixed.
Thus, changing the cluster size changes their spatial organization: smaller defect diameters correspond to a more fragmented architecture with many smaller weak clusters, whereas larger defect diameters correspond to fewer but larger weak regions.
We consider two indentation configurations: centered and off-centered.
In the centered case, a weak cluster lies beneath the indenter, allowing us to test the strongest possible coupling between the indenter and the defect.
In the off-centered case, the central region below the indenter is kept free of weak clusters, allowing us to test whether nearby defects remain visible when the indenter samples an intact load-bearing region.
The centered defect is  considered in this section while off-centered in the next.

%\subsubsection{Centered defect}
%\label{sec:cluster_defect_c}

\begin{figure}[hbt!]
    \centering
    \includegraphics[width=0.95\linewidth]{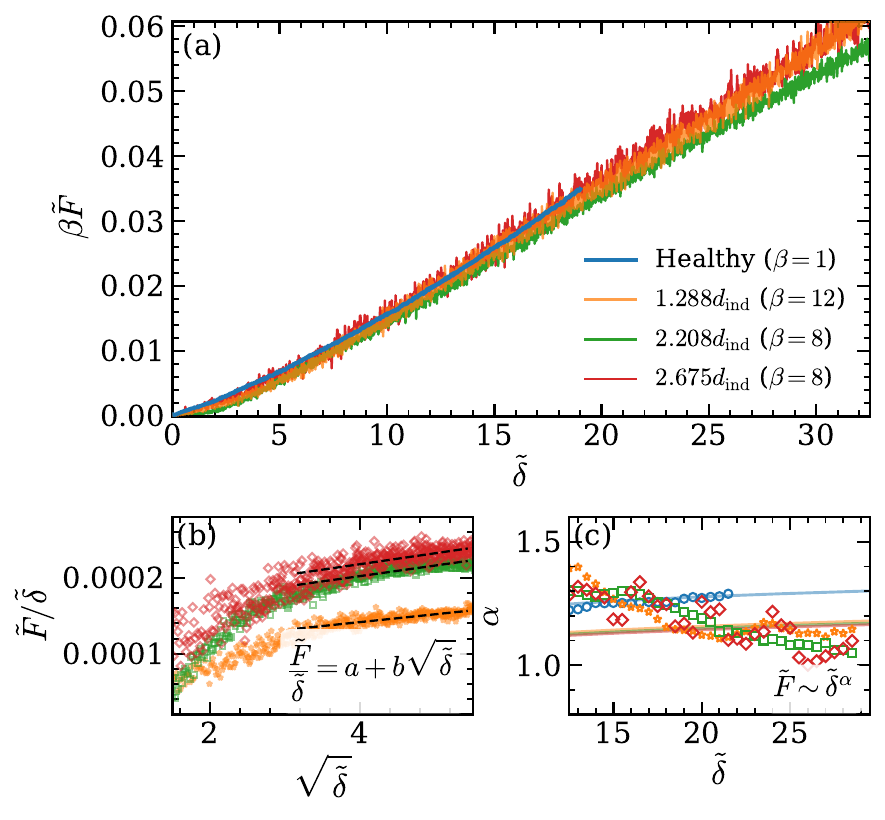}
    \caption{Indentation response of networks with centered clustered weak fibres, for three cluster diameters $d_{\rm defect}/d_{\rm ind}$, with the healthy network shown for comparison. (a) Normalised force $\tilde F = F/(pA)$ against normalised depth $\tilde\delta = \delta/d_{\rm f}$, each defective response scaled by the factor $\beta$ given in the legend. (b) The unscaled data as $\tilde F/\tilde\delta$ against $\sqrt{\tilde\delta}$; the healthy network lies above the plotted range and is not shown. Dashed lines are fits of $\tilde F/\tilde\delta = a + b\tilde\delta^{1/2}$. (c) Local exponent $\alpha$ from fits of $\tilde F \sim \tilde\delta^{\alpha}$, with the prediction of the fitted coefficients (lines). Colours as in panel (a).}
    \label{fig:indent_weak_cluster}
\end{figure}

%We now consider the networks in which weak fibers are organized into compact clusters, with one weak cluster positioned birectly beneath the indenter.
%
Three centered-cluster diameters are considered, corresponding to 
$d_{\text{defect}}/d_{\text{ind}}=$1.288, 2.208, and 2.675, all of which exceed the indenter diameter.
Fig.~\ref{fig:indent_weak_cluster}(a) shows the force--indentation response of weak networks together with the healthy network, each rescaled by the factor $\beta$ required to bring it onto the healthy curve. 
The rescaling factors, $\beta=12,8,8$, are about an order of magnitude larger than the $\beta = 1.5$ obtained for diffuse weakening at identical defect content. 
Thus, at fixed total defect content, the spatial organization of weakening governs the measured indentation resistance.
The equal factors for the two largest clusters indicate that once a cluster substantially exceeds the indenter, the force depends only weakly on its size, although cluster size and number vary together in this series.
A cluster smaller than the contact, which would leave the intact network beneath part of the probe, was not examined.

The 8–12-fold lower force of the centered configurations is not explained by a softer layer. 
The inflated profiles of these networks give in-plane moduli of 1.2–1.6 MPa, similar to the diffused one. 
With their tensions the membrane solution predicts $a_\text{N}=6.8$–$7.5\times10^{-4}$, only about 20\% below the healthy $9.12\times10^{-4}$. 
The drop arises instead because the probe rests on slack material, which decouples it from the tension of the layer.

The shape of the response changes as well. 
The data in Fig.~\ref{fig:indent_weak_cluster}(b) are not exactly linear, as it exhibits significant scatter about the fitted lines.
The local exponent $\alpha$ decreases with depth in all three clustered configurations, from $\alpha\approx1.3$ towards $\alpha\approx1.1$, whereas it increases in the healthy network (Fig.~\ref{fig:indent_weak_cluster}(c)).
Neither description can produce such a decrease: with the positive fitted $a$ and $b$, the two-term form gives an exponent that increases with depth, Eq.~\eqref{eq:alpha}, and so does the membrane solution, Eq.~\eqref{eq:alpha_m}.
The two-term form therefore does not describe a network indented directly on a weak cluster, and the coefficients fitted to these networks are only nominal (Table~\ref{tab:membrane} gives them in parentheses).

\subsubsection{Clustered weakening: off-centered defect}
\label{sec:cluster_defect_oc}

\begin{figure*}[hbt!]
  \centering
  \begin{minipage}{0.49\linewidth}
    \centering
    \includegraphics[width=\linewidth]{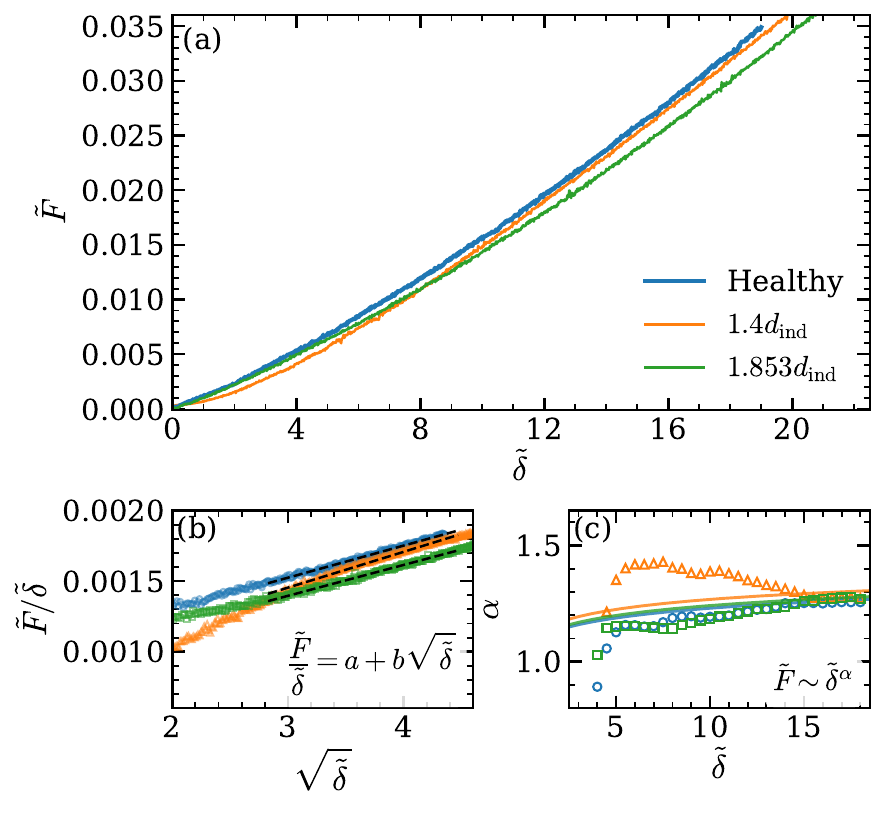}
  \end{minipage}
  \hfill
  \begin{minipage}{0.49\linewidth}
    \centering
    \includegraphics[width=\linewidth]{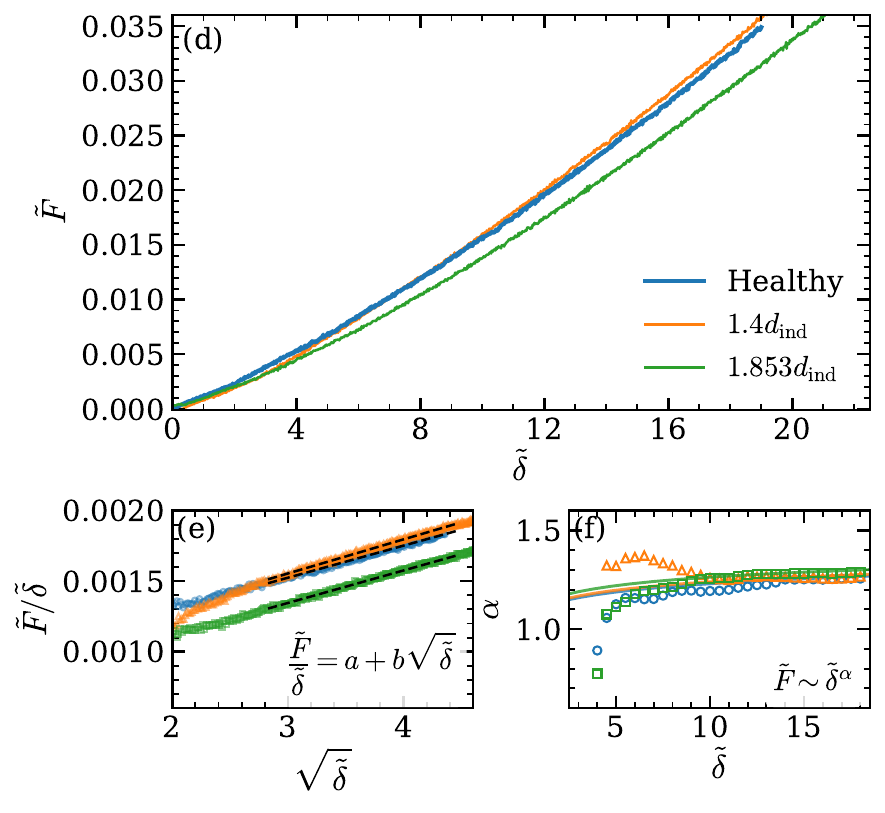}
  \end{minipage}
  
  \caption{Indentation response of networks with off-centered clustered weak fibres, for cluster diameters $d_{\rm defect}/d_{\rm ind} = 1.853$ (a--c) and $2.675$ (d--f). Legends give the avoidance diameter $d_{\rm avoid}/d_{\rm ind}$, the diameter of the region beneath the probe kept free of weakened material; the healthy network is shown for comparison. (a,d) Normalised force $\tilde F = F/(pA)$ against normalised depth $\tilde\delta = \delta/d_{\rm f}$. Dashed lines are fits of $\tilde F = a\tilde\delta + b\tilde\delta^{3/2}$. (b,e) The same data as $\tilde F/\tilde\delta$ against $\sqrt{\tilde\delta}$, on which the two-term form is a straight line of intercept $a$ and slope $b$. (c,f) Local exponent $\alpha$ from fits of $\tilde F \sim\tilde\delta^{\alpha}$ over moving windows of width $4 d_{\rm f}$ (symbols), with the prediction of the fitted coefficients (lines). Colours as in the corresponding force panel.}
  \label{fig:indent_weak_cluster_oc}
\end{figure*}

We now displace the weak clusters away from the indentation axis, keeping a central region of diameter $d_{\text{avoid}}$ beneath the indenter free of weakened material.
Two avoidance diameters, $d_{\text{avoid}}/d_{\text{ind}}=1.4$ and 
1.853, are considered for each of two cluster diameters, $d_{\text{defect}}/d_{\text{ind}}=1.853$~(Fig.~\ref{fig:indent_weak_cluster_oc}a--c) and 2.675 (Fig.~\ref{fig:indent_weak_cluster_oc}d--f). 

In contrast to the centered configurations, no rescaling is required: all off-centered responses are plotted with $\beta=1$.
Where positioning a cluster beneath the indenter suppressed the force by a factor of eight to twelve, displacing the same clusters clear of the contact restores the response to within roughly 15\% of the healthy network across all four configurations.
The mechanical signature of a localized defect is therefore governed almost entirely by whether it lies within the indenter-induced deformation field, and only weakly by its presence elsewhere in the network.

The near-healthy indentation response includes two opposite changes.
First, the weak cluster softens the layer as a whole, to in-plane moduli of 1.4--1.8 MPa against 3.62 MPa, thus carrying 20--26\% less tension.
The intercepts reflect this: in three of the four configurations, $a$ is 17–26\% below the healthy value, close to the membrane prediction $a_\text N=(7.0$–$7.5)\times10^{-4}$, and inverting it recovers the reduced modulus to within 15\% (Table~\ref{tab:membrane}).
Second, the excess stiffening is larger than in the healthy network, $b-b_{\text{N}}=(1.4-1.8)\times10^{-4}$ against $1.07\times 10^{-4}$.
Local stretching accounts for only 30–40\% of it, against about three quarters in the healthy and diffuse networks. 
Most of it arises because the slack clusters exclude load from part of the layer, so the intact backbone between them carries more tension and resists the probe more strongly.
The lower tension reduces the force and excess tension increases it. 
The two effects nearly cancel each other and the single fitted stiffness cannot distinguish these networks from the intact one.
Separating the response into the intercept, set by the global tension, and the slope, which collects the local stiffening, reveals the weakening that the total force conceals.

The local exponent retains a further signature, of the clusters rather than of the softened layer.
For the smaller avoidance diameter, $d_{\rm avoid} = 1.4 d_{\rm ind}$, $\alpha$ rises well above the healthy trend at small depths and decays towards it, reproducing in attenuated form the behaviour identified for centered clusters.
This elevation exceeds the value predicted by the two-term fit of the same curve, so it is not a consequence of the fitted coefficients but a departure from the assumed form: at small depths the response is partly accommodated by the nearby weakened material, which neither term describes.
For the larger avoidance diameter the exponent follows the healthy trend closely in both panels. 
The amplitude of the signature is therefore set by how close the nearest weakened material lies to the probe.

At fixed avoidance of 1.4$d_{\text{ind}}$, the exponent merges more rapidly with the healthy trend for the larger cluster. 
Since the total defect content is fixed, larger clusters are correspondingly fewer, so less weakened material lies within reach of the deformation field.
With two values of each geometric parameter and a single network realization per configuration, these results establish trends rather than a quantitative law.

\subsubsection{Limits of macroscopic discrimination}
\label{sec:limits_macro}

At fixed total defect content, the factor needed to rescale the defective response onto the healthy one spans roughly an order of magnitude: $\beta=1.5$ for diffuse weakening, $\beta=8-12$ when same weakened bonds are clustered with one beneath the indenter, and about 1 when the weak clusters are displaced clear of the contact.
The measured force therefore cannot be read as a measure of the amount of weakening, which has two consequences for measurements.
A cluster much larger than the indenter decouples the probe from the layer tension, so further growth leaves the response unchanged and the size of the cluster cannot be determined. 
Clusters displaced from the contact keep the response within about 15\% of the healthy one, comparable to the fluctuations of an individual curve, so a cortex with substantial localised weakening may be classified as healthy purely because of where the probe is placed.

Table~\ref{tab:membrane} summarizes the two-term coefficients of all networks together with moduli obtained by inverting the measured intercept $a$ (Eq.~\eqref{eq:Ebar_from_a}).
For healthy, diffuse, and three of the four off-centered networks, $\bar{E}_a$ lies within 2--15\% of von-K\'arm\'an modulus. 
The intercept therefore reflects the global tension whenever indenter rests on the intact network.
Over centered clusters, the inversion returns 5--20 kPa, two orders too low, because the slack cluster beneath the probe carries little tension.
A Hertz fit of the whole force-depth curve gives around 43 kPa for healthy, 28 kPa for diffuse, and 36--40 kPa for off-centered networks, which lie one to two orders of magnitude below the in-plane  modulus.
The Hertz law assumes that the force comes from deforming an unstressed elastic half-space and therefore scales with the modulus, whereas here it is set by the pre-tension of a thin stretched layer, which depends only on the cube root of the in-plane modulus (Eq.~\eqref{eq:N_analytic}).
\begin{table*}[hbt!]
\small
\caption{Two-term coefficients from the indentation simulations ($a$, $b$) and predicted by the membrane solution at constant tension ($a_\mathrm{N}$, $b_\mathrm{N}$), and with local stretching ($a_\mathrm{m}$, $b_\mathrm{m}$), all $\times 10^{-4}$. All are fitted over $2 \le \tilde\delta \le \tilde\delta_\mathrm{max}$ ($\tilde\delta_\mathrm{max} = 16$, 18, 20 and 30 for healthy, diffuse, off-centred and centred networks), with $N$ from the von K\'arm\'an fit of each inflated profile. Off-centred networks are labelled by $d_\text{defect}/d_\text{ind}$ and $d_\text{avoid}/d_\text{ind}$. In-plane modulus (MPa) from the von K\'arm\'an fit
and from Eq.~\eqref{eq:Ebar_from_a}, averaged over
$4 \le \delta_0/d_\mathrm{f} \le 16$, applied to $a$. Eq.~\eqref{eq:twoterm} does not hold for the centered networks.}
\label{tab:membrane}
\centering
\begin{tabular}{l|cccccc|cc}
\hline
 & & & & & & & \multicolumn{2}{c}{$\bar E$ (MPa)}  \\
Network & $a$ & $a_\mathrm{N}$ & $a_\mathrm{m}$  & $b$ & $b_\mathrm{N}$ & $b_\mathrm{m}$
 & von K\'arm\'an &  from $a$ \\
\hline
\rowcolor{green!15}
Healthy                    & 8.89 & 9.24 & 7.76 & 2.14 & 1.03 & 1.91 & 3.62 &  3.26   \\
Diffuse                    & 6.65 & 6.88 & 6.03 & 1.32 & 0.74 & 1.19 & 1.42 &  1.37   \\
Off-centred (1.853/1.4)    & 6.93 & 7.19 & 6.17 & 2.54 & 0.75 & 1.28 & 1.58 &  1.55   \\
Off-centred (1.853/1.853)  & 7.40 & 7.55 & 6.41 & 2.19 & 0.79 & 1.38 & 1.84 &  1.88   \\
Off-centred (2.675/1.4)    & 8.29 & 6.98 & 6.02 & 2.42 & 0.73 & 1.22 & 1.44 &  2.65   \\
Off-centred (2.675/1.853)  & 6.54 & 7.11 & 6.11 & 2.31 & 0.74 & 1.26 & 1.53 &  1.30   \\
\rowcolor{gray!15}
Centred (1.288)         & (1.00) & 7.48 & 5.89 & (0.10)   & 0.70 & 1.38 & 1.63 &  0.005 \\
\rowcolor{gray!15}
Centred (2.208)         & (1.47) & 6.82 & 5.51 & (0.14)   & 0.64 & 1.19 & 1.23 &  0.015  \\
\rowcolor{gray!15}
Centred (2.675)         & (1.60) & 7.18 & 5.72 & (0.14)   & 0.67 & 1.29 & 1.44 &  0.020  \\
\hline
\end{tabular}
\end{table*}
Not all information is lost, however. The depth-resolved exponent $\alpha = \mathrm{d}\ln F/\mathrm{d}\ln\delta$ separates cases that the force level does not.
As it remains unchanged under a constant rescaling of the force, it is insensitive to the absolute force calibration.
For centered clusters, the sign of $\mathrm{d}\alpha/\mathrm{d}\delta$ distinguishes all defect sizes examined from the healthy network. 
For off-centered clusters, $\alpha$ can retain a residual signature when weakened material lies sufficiently close to the indenter, but this signature weakens with increasing separation and can eventually become  indistinguishable from the healthy response.
Table~\ref{tab:alpha} summarizes these trends. 
Both membrane solutions predict an exponent that increases with depth, by nearly the same amount for every network, so a decreasing exponent cannot arise from the continuum response and identifies weak material beneath or close to the probe.

\begin{table}[hbt!]
\small
\caption{Local exponent $\alpha$ from moving-window fits, Eq.~\eqref{eq:alpha_def}, at the start and end of the depth range analysed for each network, and its extremum where it is not monotonic; off-centred networks are labelled by $d_\text{defect}/d_\text{ind}$ and $d_\text{avoid}/d_\text{ind}$. For comparison, between $\tilde\delta=5$ and 15 the constant-tension solution gives $\alpha_\text{N}=1.12$ to 1.14 and the local stretching solution about 1.15 to 1.22 for all networks (1.17 to 1.26 for the healthy one).}
\label{tab:alpha}
\centering
\begin{tabular}{l c c c c}
\hline
Network & $\tilde\delta$ range & start & extremum & end \\
\hline
Healthy & 5--18 & 1.14 & -- & 1.27 \\
Diffuse & 5--18 & 1.06 & -- & 1.26 \\
Off-cent.\ 1.853/1.4 & 5--18 & 1.38 & 1.45 & 1.30 \\
Off-cent.\ 1.853/1.853 & 5--18 & 1.18 & -- & 1.31 \\
Off-cent.\ 2.675/1.4 & 5--18 & 1.35 & 1.40 & 1.29 \\
Off-cent.\ 2.675/1.853 & 5--18 & 1.15 & -- & 1.32 \\
Centred (all) & 5--28 & 1.33--1.53 & -- & 1.08--1.17 \\
\hline
\end{tabular}
\end{table}

Since the above configurations contain identical amounts of compliant material yet respond differently based on the indenter placement. 
The distinction must lie in how load is transmitted through the network rather than in how much of it is weakened. 
Section~\ref{sec:micro} examines that transmission directly.

\subsection{Microscopic origin of indentation signatures}
\label{sec:micro}
To identify the microscopic origin of the macroscopic indentation signatures discussed above, we examine how load and deformation are transmitted through the network. 
We first consider the axial bond-tension field, followed by the corresponding displacement field, which is more directly accessible experimentally.

\subsubsection{Bond-tension fields}
\label{sec:tension}

\begin{figure*}[hbt!]
    \centering
    \includegraphics[width=0.95\linewidth]{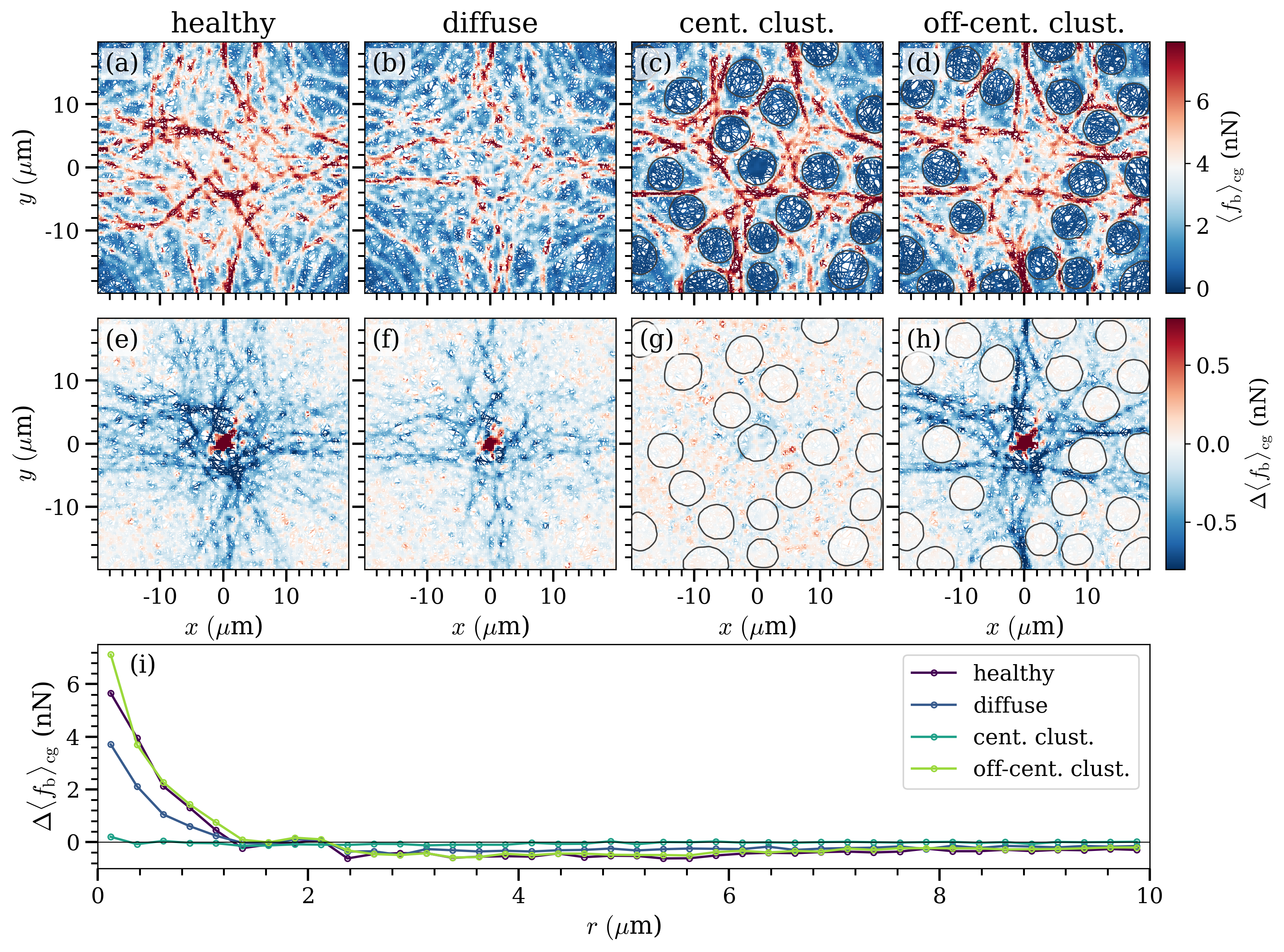}
    \caption{Coarse-grained axial bond tension
    $\langle f_{\mathrm{b}}\rangle_{\mathrm{cg}}$ in the pressure-loaded reference state (a)--(d) and its change $\Delta\langle f_{\mathrm{b}}\rangle_{\mathrm{cg}}$ at $\delta/d_f = 20$ relative to that state (e)--(h), for healthy, diffusely weakened, centred-clustered and off-centred-clustered networks. The clustered cases use $d_{\mathrm{defect}}/d_{\mathrm{ind}} = 1.853$, with an avoidance diameter $d_{\mathrm{avoid}}/d_{\mathrm{ind}} = 1.853$ in the off-centred case. The two rows use separate colour scales. Grey contours enclose regions where the weakened-bond fraction exceeds one half; the indentation axis is located at $x=y=0$. (i) Radial profiles of $\Delta\langle f_{\mathrm{b}}\rangle_{\mathrm{cg}}$, averaged over raw per-bond values in annular bins about the axis and therefore unaffected by the smoothing used for display in (a)--(h).}
    \label{fig:bond_tension}
\end{figure*}

To identify the origin of the behaviour described above we examine the axial tension carried by individual bonds.
For each bond we evaluate $f_{\mathrm{b}} = k(r - r_0)$ and average over bonds within a coarse-graining bin.
We report both the tension in the pressure-loaded reference state,
$\langle f_{\mathrm{b}}\rangle_{\mathrm{cg}}$ ($\delta=0$), and its change from initial state upon indentation, $\Delta\langle f_{\mathrm{b}}\rangle _{\mathrm{cg}}$.
The reference state shows how the load is carried before the probe touches. 
The change tests the local stretching picture of Section~\ref{sec:membrane_indentation}.
If the excess stiffening arises from local stretching, indentation should raise the tension near the contact, within a radius of order $r^*$.
Radial profiles show the mean raw per-bond tension within annular bins about the indentation axis.
Fig.~\ref{fig:bond_tension} shows both quantities for the four architectures at a common indentation depth of $\delta/d_{\mathrm{f}}=20$.
The reference state (Fig.~\ref{fig:bond_tension}(a--d)) reveals a strongly heterogeneous tension distribution. 
Even in the healthy network, the applied pressure is supported primarily by a sparse, filamentary backbone of highly tensioned bonds embedded within a weakly loaded network, as known for disordered fibre networks~\cite{head2003PRL,Beroz2017NatCom}.
This heterogeneity is therefore a property of the loaded network, present before the indenter makes contact, rather than something created by indentation.
The weakened regions are clearly resolved in the reference state. 
In both clustered networks, the outlined clusters appear as regions of near-zero tension. 
The excess contour length of their fibers makes them slack rather than merely compliant. 
The tensioned backbone passes between them. 
In the diffusely weakened network no such voids appear, instead the backbone itself is weaker and more fragmented throughout.

Turning to the change in tension upon indentation (Fig.~\ref{fig:bond_tension}(e--h)), the healthy network in Fig.~\ref{fig:bond_tension}(e) develops a compact region of increased tension (red) at and around the contact, consistent with the local stretching of the layer.
Around it, the tension is reduced along filamentary paths as the indenter flattens the pressure-inflated layer.
These paths follow the backbone already identified in the reference state.
Indentation thus redistributes load along pre-existing tension-bearing routes rather than establishing new ones.
The diffuse (Fig.~\ref{fig:bond_tension}(f)) and off-centered (Fig.~\ref{fig:bond_tension}(h)) networks show the same pattern.
The centered-clustered configuration (Fig.~\ref{fig:bond_tension}(g)) behaves differently.
A weak tension propagation is present, but its amplitude is substantially smaller and is obscured on the common color scale used for comparison here.
Rescaling the bond-tension field for this case alone (Fig.~\ref{fig:bond_tension_c}) reveals a load-bearing structure that develops progressively with depth, indicating gradual recruitment of the initially slack region and its surroundings.
Because the material beneath the contact carries little axial tension in the reference state, indentation primarily deforms the slack region and transmits only a limited axial load to the surrounding network.

The radial profile in Fig.~\ref{fig:bond_tension}(i) quantifies the tension increase at and around the contact.
At $\tilde{\delta}=$20 it is confined to $r\lesssim 1.3~\mu$m for the heathy network, comparable to $r^{\ast}\approx1.0~\mu$m (Eq.~\eqref{eq:eps_ind}).
Its magnitude follows the same order as the excess stiffening $b-b_\text N$ (Table \ref{tab:membrane}): largest for the off-centered networks ($1.4$–$1.8\times10^{-4}$), smaller for the healthy ($1.07\times10^{-4}$) and diffuse ($0.57\times10^{-4}$) networks, and negligible for the centered cluster, where the probe rests on slack material.

\subsubsection{Displacement fields}

\begin{figure*}[hbt!]
    \centering
    \includegraphics[width=0.95\linewidth]{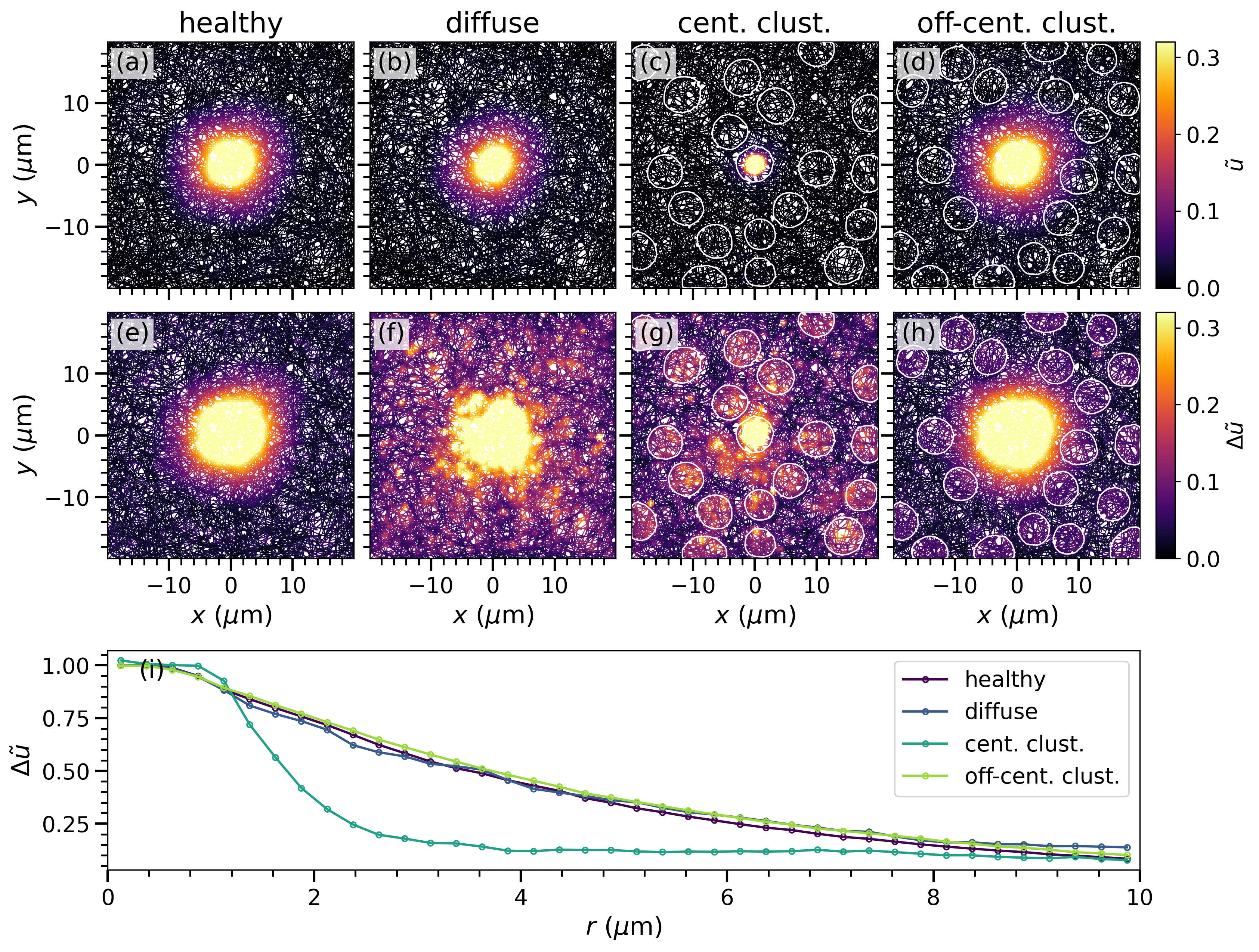}
    \caption{Normalised displacement fields for the healthy, diffusely weakened, centred-clustered, and off-centred-clustered networks at $\delta/d_f=20$. Panels (a)--(d) show the cumulative displacement $\tilde{u}=|\mathbf{u}|/\delta$, while (e)--(h) show the incremental displacement $\Delta\tilde{u}=|\Delta\mathbf{u}|/\Delta\delta$ between successive configurations. The clustered cases use $d_{\mathrm{defect}}/d_{\mathrm{ind}}=1.853$, with $d_{\mathrm{avoid}}/d_{\mathrm{ind}}=1.853$ for the off-centred case. Marked contours indicate the weakened regions, as in Fig.~\ref{fig:bond_tension}. Panel (i) shows the radially averaged incremental displacement $\Delta\tilde{u}$ about the indentation axis.}
    \label{fig:displacement}
\end{figure*}

The bond-tension fields, although very useful, are not always accessible experimentally. 
However, the displacement-fields can be more readily measured. 
It is therefore worth asking how much of the load-transfer picture of Section~\ref{sec:tension} survives in a quantity an experiment could record.
We consider two normalised fields.
First is the cumulative-field, $\tilde u = |\mathbf u|/\delta$, which gives the total displacement of a particle relative to the total indentation depth and thus describes the overall shape of the deformed region.
Second is incremental field, $\Delta\tilde u = |\Delta\mathbf u|/\Delta\delta$, evaluated between successive configurations, which gives the fraction of the imposed indentation accommodated locally at the current depth. Both equal unity for material moving rigidly with the indenter.
Thus the regions with $\Delta\tilde u \approx 1$ are dominated by the contact rather than the network.
The results are shown in Fig.~\ref{fig:displacement} for the same networks as those in Fig.~\ref{fig:bond_tension} under the same indentation depth of $\delta/d_{\text{f}}=20$.

The cumulative fields in Fig.~\ref{fig:displacement}(a--d) carry little architectural information. 
Healthy, diffusely weakened and off-centered networks produce almost indistinguishable compact, smoothly decaying deformation regions. 
Only the centered configuration, Fig.~\ref{fig:displacement}(c), is distinguishable, its deformed region confined to little more than the contact itself.

The incremental fields in Fig.~\ref{fig:displacement}(e--h) are more informative.
The healthy network shows a high-displacement region beneath the contact decaying smoothly with radius. 
The diffusely weakened network reproduces this pattern in distorted and fragmented form, with an irregular central region and isolated patches of elevated displacement scattered through the network; it is recognizably a noisier version of the healthy case rather than a distinct one.
The centered configuration shows a narrow high-displacement region at the contact together with intermediate displacement within the slack clusters themselves, including those remote from the indentation axis.
The off-centered configuration is again close to the healthy case, differing only in a slightly elevated displacement within the clusters.

The radially averaged profiles are shown in Fig.~\ref{fig:displacement}(i).
Values are near unity for $r<1~\mu$m, thus this region is contact-dominated rather than the network.
Only the centered configuration separates clearly, falling to $\Delta\tilde u \approx 0.25$ by $r \approx 2.5~\mu$m and remaining at a low plateau thereafter.
The other three decay in a similar manner and converge with the centered case only  near $r \approx 9~\mu$m.

Incremental displacement fields and radial profiles therefore clearly reveal a weak cluster beneath the probe, whose slack material confines the deformation just as it suppresses the force.
Diffuse weakening only roughens the incremental field and clusters away from the probe barely change it, so neither is distinguished from the healthy network in the radial profiles.

\section*{Discussion and Conclusions}

We combined a microstructure-resolved model of a pressure-loaded, cortex-like fiber network with continuum membrane solution and examined how the spatial organization of mechanical weakening affects indentation. 
The same amount of weakening can leave the force almost unchanged or lower it tenfold. 
Which of the two occurs depends on how the probe couples to the load-bearing network beneath it. 
A single force curve is therefore an unreliable measure of how much the cortex is weakened.

The bond-tension fields explain these macroscopic differences in behaviour. 
Even in the healthy network, load is carried by a sparse filamentary backbone rather than uniformly. 
Clustered weak regions are largely slack in the pressure-loaded state and interrupt this load-bearing structure locally. 
When such a region lies beneath the indenter, little additional axial tension develops around the contact.
When intact region beneath it, load distributes along the pre-existing backbone. 
The displacement fields retain this distinction only partially: the centered defect is clearly identifiable, while diffuse and off-centered defects remain much closer to the healthy response.

%%%%%%%%%%%%%%%% Memb Sol discussion
The membrane solution separates the force into two parts: the response of the layer at its existing tension, and the stiffening beyond it.
With tension measured independently from inflated profile, the first part captures the shallow response wherever the probe rests on the intact backbone.
The second part, the excess stiffening $b-b_\text{N}$, arises mainly from local stretching of the layer near the contact.
This is also why the two-term form fits over the depths probed. 
Near weak clusters, it arises mainly from load channelled through the intact backbone. 
Indentation therefore reports the tension of the layer and its local transmission rather than the elastic modulus directly.

Two features of the force-response still reveal the weakening. 
The intercept reflects the global tension. 
It recovers the softening of the layer even when the total force is close to the healthy one. 
The depth-dependent exponent increases with depth in every configuration except when a cluster lies beneath the probe, where it decreases. 
Its sign therefore flags weak material beneath the probe, independently of force calibration.

For indentation of cell-cortex, the constant-tension solution offers a simple reference.
Given the cortical tension, the probe radius and the cell radius $R_\text{cell}$, it predicts the force in closed form.
For a rounded cell, Laplace’s law gives $p/2N=1/R_\text{cell}$.
The curvature of the cell then takes the place of that of the inflated dome in $R^*$; in the simulations, $2N/p=56~\mu$m, close to the 50 $\mu$m radius of the dome.
The lateral size $R_\text{m}$, which has to be taken of the order of the cell size, enters only logarithmically. 
An intercept close to $a_\text N$, with an exponent that increases with depth, then indicates an intact, tensioned cortex beneath the probe.
Three effects absent from the model limit this use. 
First, adhesion between probe and membrane enlarges the contact and lowers the force at small depths, and can therefore mimic weak material beneath the probe. 
Second, the cytoplasm, nucleus and underlying cytoskeleton, represented here only by a pressure, add a resistance that depends on depth and loading rate and adds to the stiffening with depth. 
Third, the closed geometry of a cell, which at larger depth calls for the solution of pressurised shells. 
The comparison is therefore most reliable at shallow depths and slow loading, once adhesion has been excluded by comparing approach and retraction curves.
%%%%%%%%%%%%%%%%%%%%%%%%

Several other limitations should be kept in mind. 
The model is a coarse-grained, passive network; crosslinker turnover, active myosin dynamics, and fluid--structure coupling are not resolved. 
The weak fiber properties are phenomenological.
The contractility and pressure enter as effective loads representative of a mitotic cortex, thus the results describe the short-time response of a fixed architecture.
Each configuration is a single network realization indented at a fixed rate, with one pressure, one indenter size and one defect fraction.
We therefore restrict our conclusions to features whose magnitude exceeds the fluctuation amplitude of an individual force curve, and the clusters comparable or smaller than contact may behave differently.
On the continuum side, the Berger approximation replaces the non-uniform in-plane stress by its average, and the local stretching estimate is not self-consistent; a full axisymmetric F\"oppl--von K\'arm\'an solution with non-uniform in-plane stress for the present geometry would show how much of the remaining 10--12\% offset is a continuum effect.

Despite the above simplifications, the implication for indentation-based assays remains clear.
A single fitted stiffness substantially under-determines the underlying network architecture. 
The shape of the force curve, through its intercept and its exponent, carries additional information.
Indentation nevertheless remains local, so detecting localised weakening requires spatial sampling.

\section*{Author contributions}
SA carried out the conceptualization, model formulation, simulations, analysis, visualization, and preparation of the original manuscript. ON assisted in model formulation. MHM supervised the work and assisted with analysis, manuscript review, and revision.

\section*{Conflicts of interest}
The authors have no conflicts of interest to disclose.

\section*{Data availability}
The LAMMPS and post-processing scripts are available on GitHub: \url{https://github.com/shubham104084/Cortex_MD.git}

\section*{Acknowledgements}
%The authors gratefully acknowledge insightful discussion with Dr. Olivier No\"el at  Le Mans University, France.
%
The authors acknowledge Prof. Franziska Lautenschl\"ager and Lucina Kainka at Saarland University for initial discussions that helped motivate the present study.

%%%END OF MAIN TEXT%%%

\section*{Appendix A}

\setcounter{subsection}{0}
\renewcommand{\thesubsection}{A.\arabic{subsection}}

\setcounter{figure}{0}
\renewcommand{\thefigure}{A\arabic{figure}}

\subsection{Analytical solution in linear approximation}
\label{app:continuum_sheet}

On the domain $\Omega = [0,L] \times [0,L] \times [0,h]$, 
\begin{comment}
static linear elasticity with no body force is governed by the Navier--Cauchy equation
%
\begin{equation}
  \mu\,\nabla^{2}\mathbf{u}
  + (\lambda+\mu)\,\boldsymbol{\nabla}
    \left(\boldsymbol{\nabla}\cdot\mathbf{u}\right) = \mathbf{0} .
\end{equation}
%
Rather than solving it directly, 
\end{comment}
we obtain the displacement field $\mathbf{u}$ by minimising the
potential energy,
\begin{align}
  \Pi & = \frac{1}{2} \int_\Omega C_{ijkl}\,
          \partial_j u_i\,\partial_l u_k \,\mathrm{d}V
          - \int_{z=0} p\,u_z \,\mathrm{d}A ,
\end{align}
where the elastic tensor $\mathsf{C}$ is assumed to obey the isotropy
condition $C_{1212} = (C_{1111}-C_{1122})/2$.

In our simulation cell, we have the boundary conditions
\[
\mathbf{u} = \mathbf{0}
    \quad \text{at} \quad x = 0,\ L \quad \text{and} \quad y = 0,\ L ,
\]
while the bottom and top walls satisfy
\begin{align}
  \sigma_{zz} = -p , \quad \sigma_{xz} &= \sigma_{yz} = 0
    & &\text{at} \quad z = 0 , \\
  \sigma_{zz} = \phantom{-}0 , \quad \sigma_{xz} &= \sigma_{yz} = 0
    & &\text{at} \quad z = h .
\end{align}
The latter two are the natural boundary conditions of $\Pi$ and are therefore
obeyed at the minimum without being imposed; only the clamping has to be built
into the ansatz.  Given the clamped boundary conditions on the side walls, the
solution can be expanded as
\[
u_i(\mathbf{r}) = \sum_{m,n} \tilde{u}_{i,mn}(z)\,
  \sin(q_m x)\sin(q_n y) \quad \text{with} \quad q_m = m\pi/L .
\]

Abbreviating $S_m(x)=\sin(q_m x)$ and $K_m(x)=\cos(q_m x)$, the gradients
entering $\Pi$ are
\begin{align}
  \partial_j u_i & = \sum_{m,n}
    \tilde{g}^{(j)}_{i,mn}(z)\,F^{(j)}_{mn}(x,y) ,
\end{align}
with
\begin{align}
  \tilde{g}^{(1)}_{i,mn} = q_m\,\tilde{u}_{i,mn} , & \quad
    F^{(1)}_{mn} = K_m S_n , \nonumber\\
  \tilde{g}^{(2)}_{i,mn} = q_n\,\tilde{u}_{i,mn} , & \quad
    F^{(2)}_{mn} = S_m K_n , \nonumber\\
  \tilde{g}^{(3)}_{i,mn} = \partial_z \tilde{u}_{i,mn} , & \quad
    F^{(3)}_{mn} = S_m S_n .
\end{align}
Only first derivatives occur, so that the energy requires the in-plane overlaps
\begin{align}
  W^{(jl)}_{mn,m'n'} & \equiv
    \int_0^L\!\!\int_0^L F^{(j)}_{mn} F^{(l)}_{m'n'}\,
    \mathrm{d}x\,\mathrm{d}y \nonumber\\
  & = \begin{cases}
    \left(\tfrac{L}{2}\right)^2 \delta_{mm'}\delta_{nn'} , & j = l \\[2pt]
    M_{m'm}\,M_{nn'} , & (jl) = (12) \\[2pt]
    M_{m'm}\,\tfrac{L}{2}\,\delta_{nn'} , & (jl) = (13) \\[2pt]
    \tfrac{L}{2}\,\delta_{mm'}\,M_{n'n} , & (jl) = (23)
  \end{cases}
\end{align}
and the combinations with $j>l$ follow by exchanging $(jl)\to(lj)$ together
with $(mn)\leftrightarrow(m'n')$.  Here
\begin{equation}
M_{mn} \equiv
  \int_0^L S_{m}(x)\,K_{n}(x)\,\mathrm{d}x = \frac{2L}{\pi}\,
    \frac{m}{m^2-n^2} \quad \text{for } m+n \text{ odd},
\end{equation}
and zero otherwise.  The contributions with $j=l$ are diagonalised by the
orthogonality of the basis functions, whereas those with $j\neq l$ carry a
cosine in at least one direction and hence couple different $\mathbf{q}$
through $M$.  With these overlaps the energy becomes
\begin{align}
  \Pi & = \frac{1}{2} \int_0^h \!\! \mathrm{d}z
      \sum_{\substack{m,n \\ m',n'}} C_{ijkl}\,
      \tilde{g}^{(j)}_{i,mn}\,\tilde{g}^{(l)}_{k,m'n'}\,
      W^{(jl)}_{mn,m'n'} \nonumber\\
  & \phantom{=} - \sum_{m,n} f_{mn}\,\tilde{u}_{3,mn}(0) ,
  \qquad
  f_{mn} = \frac{4pL^2}{\pi^2 mn} ,
\end{align}
the load coefficient being non-zero for $m$ and $n$ both odd, so that only the
odd--odd modes are driven by the applied pressure.

What remains is a one-dimensional integral in $z$, which we discretise on the
equidistant grid $z_s = s\,\Delta z$ with $\Delta z = h/n_z$ and
$s = 0,\dots,n_z$, the unknowns being the nodal amplitudes
$\tilde{u}_{k,mn,s} \equiv \tilde{u}_{k,mn}(z_s)$.  Evaluating each interval at
its midpoint,
\begin{align}
  \partial_z \tilde{u}_{k,mn} \big|_{s+\frac{1}{2}}
   & = \frac{\tilde{u}_{k,mn,s+1}-\tilde{u}_{k,mn,s}}{\Delta z} ,
  \\
  \tilde{u}_{k,mn} \big|_{s+\frac{1}{2}}
   & = \frac{\tilde{u}_{k,mn,s+1}+\tilde{u}_{k,mn,s}}{2} ,
\end{align}
turns $\Pi$ into a quadratic form in the $3 N_q (n_z+1)$ nodal amplitudes,
where $N_q$ denotes the number of retained $\mathbf{q}$ vectors.  Since only
first derivatives appear, no values outside $0 \le s \le n_z$ are required.

The generalised force conjugate to a coefficient is
\begin{equation}
  \mathcal{F}_{k,mn,s} = - \frac{\partial \Pi}{\partial \tilde{u}_{k,mn,s}}
    = f_{mn}\,\delta_{k3}\,\delta_{s0}
      - \left( \mathsf{A}\,\tilde{\mathbf{u}} \right)_{k,mn,s} ,
\end{equation}
where $\mathsf{A}$ collects the elastic contributions.  Equilibrium is reached
when all $\mathcal{F}_{k,mn,s}$ vanish, which we solve iteratively.  As
$\mathsf{A}$ derives from an energy it is symmetric and positive definite, so
that conjugate gradients may be used.

Since $M_{mn}$, the $q_m$ and $C_{ijkl}$ are all independent of $z$, each block
of $\mathsf{A}$ is a Kronecker product of $M$ acting in $x$, $M$ acting in $y$,
and a banded matrix acting in $z$.  The matrix therefore never has to be
assembled: it can be applied factor by factor, which reduces the memory demand
from $\mathcal{O}(N_q^2 n_z^2)$ to $\mathcal{O}(N_q n_z)$ and makes the
iterative solution of the coupled system practical.

%\subsection{Nonlinear thin-sheet approximation}

\subsection{Finite difference solution of the pressure-loaded profile}
\label{app:continuum}

The Eq.~\eqref{eq:berger} is discretized on a uniform $n\times n$ grids of interior nodes with spacing $\Delta = L/(n+1)$, there $L$ is the side length of free patch.
With $w_{i,j}$ the deflection at each node $(i,j)$, the discrete system reads:
\begin{equation}
\left( D\,\mathbf{B} - N\,\boldsymbol{\Lambda} \right) \mathbf{w} = p\,\mathbf{1},
\label{eq:fd_system}
\end{equation}
where $\mathbf{w}$ collects the nodal deflections, $\mathbf{1}$ is a
vector of ones, and $\boldsymbol{\Lambda}$ and $\mathbf{B}$ are the
discrete Laplacian and biharmonic operators. The Laplacian is
represented by the standard five-point stencil,
\begin{equation}
(\boldsymbol{\Lambda}\mathbf{w})_{i,j} = \frac{1}{\Delta^{2}}
\left( w_{i+1,j} + w_{i-1,j} + w_{i,j+1} + w_{i,j-1} - 4\,w_{i,j} \right),
\label{eq:fd_laplacian}
\end{equation}
and the biharmonic operator,
$\nabla^4 w = w_{xxxx} + 2w_{xxyy} + w_{yyyy}$, by the thirteen-point
stencil
\begin{equation}
\begin{split}
(\mathbf{B}\mathbf{w})_{i,j} = \frac{1}{\Delta^{4}} \Big[\,
 & 20\,w_{i,j}
 - 8 \left( w_{i+1,j} + w_{i-1,j} + w_{i,j+1} + w_{i,j-1} \right) \\
 & + 2 \left( w_{i+1,j+1} + w_{i+1,j-1} + w_{i-1,j+1} + w_{i-1,j-1} \right) \\
 & + w_{i+2,j} + w_{i-2,j} + w_{i,j+2} + w_{i,j-2} \,\Big].
\end{split}
\label{eq:fd_biharmonic}
\end{equation}
The displacement is set to zero along all four edges, $w_{0,j} = w_{n+1,j} = w_{i,0} = w_{i,n+1} = 0$.
The second boundary condition of the fourth-order operator is imposed through ghost points built into $\mathbf{B}$: \(w_{-1}=w_1\) for clamped edges with zero normal slope, and  \(w_{-1}=-w_1\) for simply supported edges with zero normal curvature.
Starting from $N = 0$, the sparse system is solved, $N$ is re-evaluated from the resulting deflection field, and the update is under-relaxed,
$N^{(k+1)} = (1-\omega)\,N^{(k)} + \omega\,N^{(k)}_{\mathrm{new}}$
with $\omega = 0.5$, until the relative change in $N$ falls below
$10^{-10}$. 
With $N$ held at zero, the scheme reproduces the linear plate deflections $w_{\max} = 0.00406\,pL^4/D$ for simply supported and $0.00126\,pL^4/D$ for clamped edges to within 0.1 and 0.4 percent, respectively; in the stretching-dominated limit it recovers $w_{\max} \propto \left( pL^4/\bar{E}h \right)^{1/3}$. 
All results use $n = 101$, at which $w_{\max}$ is converged to within 0.1 percent.
For both the continuum descriptions, the effective modulus $\bar{E}$ is obtained by minimizing the root-mean-square deviation between the calculated and coarse-grained network over the same fitting interval.

\subsection{Evolution of bond-tension redistribution in the centred-cluster configuration}
\label{app:bond_tension_c}

\begin{figure*}[hbt!]
    \centering
    \includegraphics[width=0.95\linewidth]{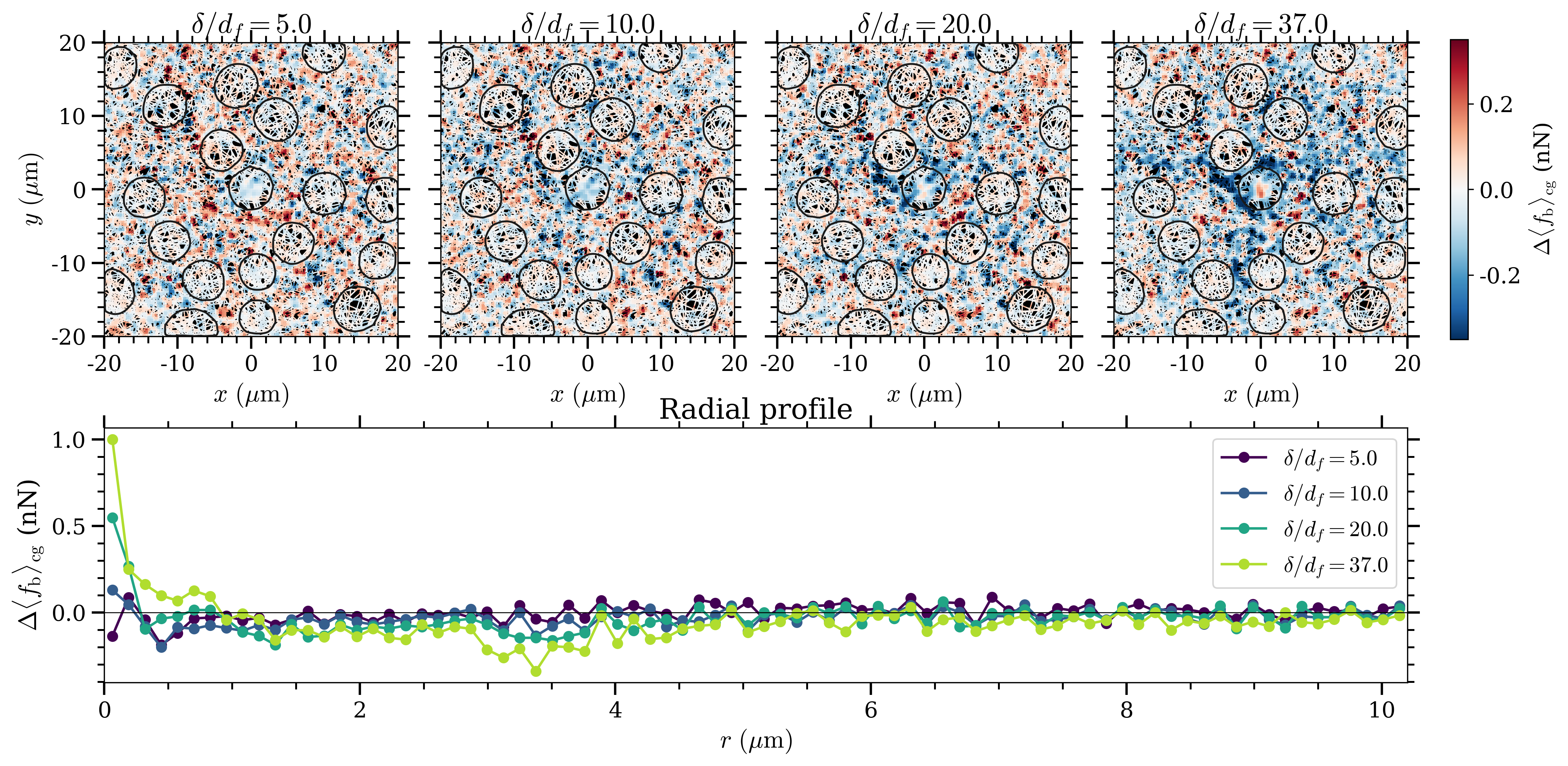}
    \caption{The top four panels shows the evolution of coarse-grained $\Delta\langle f_{\mathrm{b}}\rangle_{\mathrm{cg}}$ with indentation depth $\delta/d_f$ for clustered-centered defect. Corresponding radial profiles of $\Delta\langle f_{\mathrm{b}}\rangle_{\mathrm{cg}}$, are shown in bottom plot.}
    \label{fig:bond_tension_c}
\end{figure*}

Fig.~\ref{fig:bond_tension_c} shows the change in coarse-grained axial bond tension for the centred-cluster configuration at increasing indentation depths, plotted using a rescaled colour range to resolve the comparatively weak response that is largely obscured on the common scale used in Fig.~\ref{fig:bond_tension}. The maps reveal progressive development of the load-bearing structure with indentation, consistent with gradual recruitment of the initially slack weakened region and the surrounding network. The corresponding radial profiles show how the magnitude and spatial extent of this redistribution evolve with indentation depth.

\balance


%%%REFERENCES%%%
\bibliography{ref} 
%You need to replace "rsc" on this line with the name of your .bib file
\bibliographystyle{rsc} 
%the RSC's .bst file

\end{document}